\documentclass[prd,aps,twocolumn,a4paper,showkeys,nofootinbib,superscriptaddress]{revtex4-2}

\usepackage{graphicx}
\usepackage{dcolumn}
\usepackage{bm}
\usepackage{hyperref}
\usepackage[mathlines]{lineno}
\usepackage{amsmath,amssymb,euscript}
\usepackage{calrsfs}
\allowdisplaybreaks
\usepackage{booktabs} 
\usepackage{tikz}
\usetikzlibrary{calc,decorations.markings}
\usetikzlibrary{snakes}
\usepackage{rotating}
\usepackage{comment}
\usepackage{empheq}
\usepackage{ulem}
\usepackage{overpic}
\usepackage{hyperref}

\newcommand{\nn}{\nonumber}
\newcommand{\R}{\mathbb{R}}
\newcommand{\bbC}{\mathbb{C}}
\newcommand{\tr}{\mbox{Tr}\;}
\newcommand{\CN}{\mathcal{N}}
\newcommand{\CH}{\mathcal{H}}

\newcommand{\CO}{\mathcal{O}}

\newcommand{\CP}{\mathbb{C}\rm P}
\newcommand{\mCH}{\mathbb{C}\rm H}
\newcommand{\spa}{\ , \ \ }
\newcommand{\ds}{\displaystyle}

\begin{document}

\title{Twisted BPS decoupling limits and AdS/CFT}
\author{Troels Harmark}
 \affiliation{Center of Gravity, Niels Bohr Institute, Copenhagen University,\\ Blegdamsvej 17, DK-2100 Copenhagen \O{}, Denmark}
  \affiliation{Nordita, KTH Royal Institute of Technology and Stockholm University,\\
  Hannes Alfv\'ens v\"ag 12, SE-106 91 Stockholm, Sweden}

\email{harmark@nbi.ku.dk}

\begin{abstract}
BPS decoupling limits of string and M-theory have led to profound insights into gravity and non-perturbative quantum theory, most notably with the AdS/CFT correspondence. 
In this paper, we initiate the study of twisted BPS decoupling limits, in which the relevant BPS states carry momentum along circle fibers geometrically twisted over a base space.
Prime examples are found in type IIB string theory on AdS$_5\times S^5$, for which the twisted limits involve the BPS states of the global symmetry algebra  PSU$(2,2|4)$.
In general, string fluctuations in a twisted BPS decoupling limit are described by non-relativistic sigma-models on emergent non-Lorentzian target spaces with a natural symplectic structure. This  further supports the broader connection between BPS limits and non-Lorentzian geometry.
We introduce analogous limits of D-branes as well, probing the same symplectic structure as the string.  
For AdS$_5\times S^5$, twisted BPS decoupling limits are dual to Spin Matrix theory limits of $\CN=4$ supersymmetric Yang-Mills theory, providing tractable and explicit quantum-mechanical descriptions of non-perturbative regimes of AdS/CFT.
\end{abstract}

\maketitle

\tableofcontents

\section{Introduction}

BPS decoupling limits in string and M-theory have proven to be an exceptionally fruitful tool for understanding holography and non-perturbative quantum physics. They isolate the dynamics in the vicinity of a supersymmetric BPS state by decoupling it from modes with finite energy above the BPS bound. 
In this way, otherwise inaccessible regimes of the theory can become tractable.

A landmark example of this is the AdS/CFT correspondence between $\CN=4$ supersymmetric Yang-Mills theory (SYM) and type IIB string theory on AdS$_5\times S^5$ \cite{Maldacena:1997re}. 
This correspondence was originally uncovered by taking a decoupling limit that isolates the near-BPS dynamics of a $1/2$-BPS configuration of parallel coincident D3-branes in flat space. Other $1/2$-BPS brane configurations likewise give rise to holographic correspondences, including D$p$-branes with $0\leq p\leq 5$ \cite{Itzhaki:1998dd}.

More generally, a large class of BPS decoupling limits are dual to discrete light-cone quantization (DLCQ) limits \cite{Sen:1997we,Seiberg:1997ad,Blair:2023noj,futureUdual}. In particular, $1/2$-BPS brane configurations in flat space can be U-dualized to a frame in which the brane charge is replaced by momentum along a circle, with the corresponding BPS state describing a null worldline. The target-space geometry of strings in the DLCQ limit can be understood by passing to a T-dual frame, where the limit becomes a non-relativistic string (NRST) limit 
thereby revealing the non-Lorentzian nature of the target space \cite{Harmark:2017rpg,Bergshoeff:2018yvt,Harmark:2018cdl,Gomis:2019zyu,Harmark:2019upf,Bergshoeff:2019pij,Bergshoeff:2021bmc,Bidussi:2021ujm}. 
Moreover, a large class of BPS decoupling limits of brane configurations preserving less supersymmetry can be U-dualized to multiple simultaneous DLCQ limits \cite{Polchinski:1996na,Gauntlett:1997cv,futureUdual}. These limits likewise exhibit an underlying non-Lorentzian geometry \cite{Blair:2023noj,Blair:2024aqz,Lambert:2024ncn,Blair:2025prd,futureUdual}.

A central point of this paper, however, 
is that not all BPS decoupling limits are U-dual to DLCQ limits. 
To see this concretely, consider the BPS bound $E \geq J_1+J_2$ in the AdS/CFT correspondence. 
For type IIB string theory on  AdS$_5\times S^5$ in global coordinates,  $E$ is the AdS$_5$ energy while $J_1$ and $J_2$ are two commuting angular momenta  of $S^5$. Thus, $J_1+J_2$ is a momentum along a specific angular direction while the BPS state $E=J_1+J_2$ corresponds to a null worldline. Crucially, 
this angular direction is a circle fiber that is geometrically twisted, {\sl i.e.}~it has a non-flat U$(1)$ connection, over a base space in a fiber bundle. 
This geometric twist breaks the local ISO$(1,1)$ symmetry and ties the scaling of the fiber to that of the base space, in contrast to the DLCQ case.

On the $\CN=4$ SYM side, the same bound $E\geq J_1+J_2$, now with $J_1$ and $J_2$ being commuting R-charges, admits a natural BPS decoupling limit. This is a Spin Matrix theory (SMT) limit in which the 't Hooft coupling is sent to zero while the energy above the BPS bound is appropriately rescaled, thereby isolating the near-BPS sector \cite{Harmark:2014mpa}. In the planar and semiclassical regime, the resulting dynamics is described by the Landau-Lifshitz sigma model \cite{Fradkin1991,Kruczenski:2003gt}, which can itself be viewed as a type of non-relativistic string theory, but of a different class from the non-relativistic string theory arising from DLCQ \cite{Harmark:2008gm,Harmark:2017rpg,Harmark:2018cdl,Harmark:2019upf,Harmark:2020vll,Bidussi:2023rfs}. This further motivates the question of what the corresponding BPS decoupling limit is on the string-theory side.

More generally, all the BPS bounds associated with the PSU$(2,2|4)$ global symmetry of type IIB string theory on  AdS$_5\times S^5$ involve geometrically twisted fiber bundles, except for one degenerate case without near-BPS fluctuations. In each case, BPS states are associated with null worldlines. Furthermore, all these BPS bounds have associated BPS decoupling limits in the dual $\CN=4$ SYM in the form of SMT limits  \cite{Harmark:2014mpa,Harmark:2019zkn,Baiguera:2020jgy,Baiguera:2020mgk,Baiguera:2021hky,Baiguera:2022pll}.%
\footnote{See also \cite{Harmark:2006di,Harmark:2006ta,Harmark:2006ie,Harmark:2007px,Harmark:2008gm}
for the initial exploration of the SMT BPS decoupling limits.} The decoupled theories, known as Spin Matrix theories (SMTs), reduce to  non-relativistic sigma-model descriptions in the planar and semi-classical regime \cite{Kruczenski:2003gt,Harmark:2008gm,Hernandez:2004uw,Stefanski:2004cw,Bellucci:2004qr,Bellucci:2006bv}.

In this paper, we show that for a BPS state in type II string theory that carries momentum along a circle fiber geometrically twisted over a base space, there exists a unique and well-defined BPS decoupling limit. We call this a {\sl twisted BPS decoupling limit}. This construction includes, in particular, all BPS states associated with the PSU$(2,2|4)$ global symmetry of AdS$_5\times S^5$. For a string probe, we analyze the limit in a T-dual frame, where the momentum mode is mapped to a winding mode, and show that the Nambu-Goto action admits a finite limit. The resulting non-relativistic sigma model 
has a natural symplectic structure, allowing the string action to be interpreted as a phase-space action \cite{Harmark:2020vll,Bidussi:2023rfs}.
In a fully covariant formulation, this structure is embedded in a non-Lorentzian target-space geometry known as U$(1)$-Galilean geometry \cite{Harmark:2017rpg}.

The string duals of the SMT limits of $\CN=4$ SYM were previously explored in \cite{Harmark:2017rpg,Harmark:2018cdl,Harmark:2019upf,Harmark:2020vll,Bidussi:2023rfs}, leading to a general two-step limiting procedure. 
In the absence of a geometric twist, the first step is a DLCQ limit, while the second involves a rescaling of time. In \cite{Bidussi:2023rfs}, these were interpreted in a T-dual frame as two successive non-relativistic limits of a string probe: first an NRST limit, T-dual to the DLCQ limit, and subsequently a non-relativistic worldsheet limit. 
However, in this paper we point out that in the presence of a geometric twist, such as one encounters for BPS states in AdS$_5\times S^5$, the first step is neither a DLCQ limit nor a BPS decoupling limit, since it requires holding fixed a quantity that is not a compact charge.  
Consequently, this put in question whether the intermediate theory, by itself, defines a consistent decoupled corner of string theory. 
Nevertheless, the combined effect of the two steps is equivalent to the twisted BPS decoupling limit introduced in this paper, in particular since the above-mentioned fixed quantity becomes a compact charge after performing the second limit. 
Thus, in the presence of the geometric twist, the relevant BPS limit is the twisted BPS decoupling limit, in line with our central point that one cannot take a DLCQ limit in this case.




Our identification of the twisted BPS decoupling limits as the holographic duals of SMT limits also enables us to study the corresponding limits for D-brane probes in AdS$_5\times S^5$. In this case, the relevant BPS states are giant gravitons, and the twisted BPS decoupling limit isolates their near-BPS fluctuations \cite{Harmark:2016cjq}. Remarkably, we find that D-brane probes experience the same U$(1)$-Galilean geometry, and hence the same underlying symplectic structure, as string probes. The near-BPS fluctuations, however, reside entirely in the off-diagonal matrix degrees of freedom, requiring the non-Abelian DBI action as the starting point for taking the limit. As a consequence, these fluctuations probe a matrix-valued generalization of U$(1)$-Galilean geometry.

Importantly, the non-relativistic sigma-models found from the twisted BPS decoupling limits of type IIB string theory on AdS$_5\times S^5$ matches those found directly from the dual side as the planar and semi-classical regime of Spin Matrix theory \cite{Harmark:2008gm,Harmark:2017rpg,Harmark:2018cdl,Harmark:2020vll}.
This shows the SMT limits provide a window into strongly coupled physics despite being defined at small 't Hooft coupling, $\lambda\ll1$ \cite{Harmark:2008gm}. This is consistent with the fact that both the 't Hooft coupling and its dual manifestation as the string tension vanish in the limit. The matching of sigma-models can be argued to be possible since the AdS$_5\times S^5$ background is protected, since we are studying fluctuations close to a BPS state, and since the match is in a semi-classical regime \cite{Harmark:2008gm}.
This suggests that twisted BPS decoupling limits may furnish a tractable corner of AdS/CFT, with the SMTs as explicit quantum-mechanical theories that describe the dynamics in full.
Non-trivial checks on the consistency of the SMT limits of $\CN=4$ SYM were performed in \cite{Harmark:2019zkn,Baiguera:2020jgy,Baiguera:2020mgk,Baiguera:2021hky,Baiguera:2022pll,Auzzi:2026gan}, demonstrating that taking the BPS decoupling limits commutes with quantisation of the theories. 
In particular, this was demonstrated for the SU$(1,1)$ SMT in \cite{Harmark:2019zkn}, by showing that the 
 SU$(1,1)$ sector of the fully quantized one-loop dilatation operator follows from a simple normal-ordering effect in SU$(1,1)$ SMT.
Moreover, the twisted BPS decoupling limits of D-branes in AdS$_5\times S^5$, together with their matching to the corresponding limits of $\CN=4$ SYM \cite{Harmark:2016cjq}, demonstrate that this framework extends beyond the planar regime.

This paper is organized as follows. In Section \ref{sec:no_twist} we set the stage by briefly reviewing DLCQ limits in type II string theory. 
In Section  \ref{sec:twistedlimit} the first example of a BPS state involving a geometrically twisted circle fiber is considered, and it is shown how this example arise in the AdS/CFT correspondence.
In Section \ref{sec:genlimit} we consider a more general class of twisted BPS decoupling limits for a string probe, revealing the natural symplectic structure of the target space, and how this is embedded in a U$(1)$-Galilean geometry. 
In Section \ref{sec:examples} we exhibit all the possible twisted BPS decoupling limits of string probes on AdS$_5\times S^5$.
In Section \ref{sec:Dbranes} we demonstrate that one can also take twisted BPS decoupling limits for D-brane probes on AdS$_5\times S^5$. For a single D-brane, one gets a BPS configuration, as shown in Section \ref{sec:single_Dbrane}. This has an alternative formulation exhibited in \ref{sec:alt_form}. For multiple D-branes, one has near-BPS fluctuations as well, and we find the decoupled matrix action describing this by taking the twisted BPS decoupling limit of the non-abelian DBI action in Section \ref{sec:multiple_Dbranes}.

\section{Untwisted limit}
\label{sec:no_twist}

We begin by briefly reviewing the discrete light-cone quantization (DLCQ) limit in string theory%
\footnote{A similar presentation with more details is found in \cite{Harmark:2025ikv}. The reviews \cite{Polchinski:1999br,Taylor:2001vb,Demulder:2023bux,Oling:2022fft} have a slightly different approach, using a Lorentz boost.
Note that one can alternatively consider the DLCQ limit of M-theory. However, the DLCQ limit of string theory is more directly relevant in this paper.}
 in order to set the stage for the novel BPS decoupling limits that we explore in the rest of this paper. 
To this end, we consider 
type II string theory in ten-dimensional flat space with a circle direction
\begin{equation}
\label{circle1}
\frac{ds^2}{\alpha'}  = R^2 ( -dt^2 + d\gamma^2 ) + dx^i dx^i \,,
\end{equation}
where $t$ is time, $\gamma$ is periodic with period $2\pi$ and $\sqrt{\alpha'}R$ is the radius of the circle. 
For later convenience time is measured in units of $\sqrt{\alpha'}R$ and we introduce the string length $\sqrt{\alpha'}$ making $R$ dimensionless.
With a non-zero momentum $p = w > 0$ along $\gamma$, $w$ being an integer, one has the BPS condition for the energy $E \geq w$ which is saturated by the $1/2$ BPS state $E=w$. Here $p$ and $E$ are measured in units of $(\sqrt{\alpha'}R)^{-1}$. 
This BPS state corresponds to the null worldline with $\gamma-t$ and $x^i$ held fixed.

In general, BPS decoupling limits zoom in on the dynamics of states with energies infinitesimally close to the BPS bound. 
For the momentum mode,
this can be implemented with the coordinate change
$t = T/\epsilon^2$ and $\gamma=\hat{\gamma}+ T/\epsilon^2$ since the energy associated with $T$ is then $E_T=(E - w)/\epsilon^2$ while the momentum is unchanged. 
The BPS decoupling limit that considers small excitations above $E=w$ 
should thus have $\epsilon \rightarrow 0$. 
Note also that in these coordinates the BPS state corresponds to the null worldline parametrized by $T$ with $\hat{\gamma}$ and $x^i$ constant.

To find the scaling of the radius $R$ we consider a string probe. Assuming a small string coupling, one finds that the energy spectrum of $E_T$ for the string is finite provided $R \propto \epsilon$ for $\epsilon \rightarrow 0$. Choosing for simplicity $\epsilon=R$ we have
\begin{equation}
\label{circle2}
\frac{ds^2}{\alpha'}  =  2 dT \, d\hat{\gamma} + \epsilon^2 d\hat{\gamma}^2  + dx^i dx^i \,,
\end{equation}
which means for $\epsilon \rightarrow 0$ the $\hat{\gamma}$ direction is null. 
The $\epsilon \rightarrow 0$ limit defined this way is known as a DLCQ limit, since the $\hat{\gamma}$ direction becomes a compact null direction in the limit.

The compact null direction would seemingly mean that the space-time geometry \eqref{circle2} is problematic. However, as realized in \cite{Gomis:2000bd,Danielsson:2000gi, Harmark:2017rpg,Bergshoeff:2018yvt}, this can be resolved by going to a T-dual frame.
Indeed, performing a spatial T-duality along $\hat{\gamma}$ on the finite-$\epsilon$ background \eqref{circle2}, one finds using the Buscher rules the following metric and Kalb-Ramond field
\begin{equation}
\label{circle3}
\frac{ds^2}{\alpha'}  =  \frac{1}{\epsilon^2} (- dT^2 + dv^2)   + dx^i dx^i
\spa 
\frac{B}{\alpha'} = - \frac{1}{\epsilon^2} dT \wedge dv \,,
\end{equation}
where the T-dual direction $v$ is periodic with period $2\pi$. 
In this T-dual frame, the single string excitations close to BPS correspond to the modes of a string with winding number $w$ along the $v$ direction in the $\epsilon \rightarrow 0$ limit. 
The $\epsilon \rightarrow 0$ limit is known as the non-relativistic string limit \cite{Gomis:2000bd,Danielsson:2000gi, Harmark:2017rpg,Bergshoeff:2018yvt}.%
\footnote{As argued in \cite{Blair:2024aqz,Harmark:2025ikv}, this is equivalent to the matrix string limit \cite{Motl:1997th,Dijkgraaf:1997vv}.} 
The Nambu-Goto action for the string is finite in this limit, and reveals that the target space is now a non-Lorentzian geometry, in the form of a torsional string Newton-Cartan geometry \cite{Harmark:2017rpg,Bergshoeff:2018yvt,Harmark:2018cdl,Gomis:2019zyu,Harmark:2019upf,Bergshoeff:2019pij,Bergshoeff:2021bmc,Bidussi:2021ujm}. 
This resolves the space-time geometry in the DLCQ limit. 

One can also consider the DLCQ limit in curved space. In the T-dual frame, this is dual to the non-relativistic string limit giving a more general torsional string Newton-Cartan geometry \cite{Harmark:2017rpg,Bergshoeff:2018yvt,Harmark:2018cdl,Gomis:2019zyu,Harmark:2019upf,Bergshoeff:2019pij,Bergshoeff:2021bmc,Bidussi:2021ujm}. In \cite{Harmark:2025ikv} the curved space DLCQ argument and the associated T-duality are examined in detail. In order for the metric to have Killing vector fields of an ISO$(1,1)$ symmetry one should in particular have the boost symmetry $\gamma \partial_t + t\partial_\gamma$. 
This requires the circle direction to be orthogonal to the transverse directions. As we shall see below, breaking this boost symmetry indeed moves one away from the DLCQ limit.

\section{Twisted limit}
\label{sec:twistedlimit}

We now initiate our study of BPS decoupling limits associated with null worldlines that involves a non-trivial fiber bundle geometry. In this section we shall examine in detail a particular example that arises for type IIB string theory on AdS$_5 \times S^5$. 

To begin, consider a space-time geometry with metric
\begin{equation}
\label{met_onlyA}
\frac{ds^2}{\alpha'} =  R^2 \Big( -dt^2 + (d\gamma+A)^2  
+ h_{ij} dx^i dx^j
\Big) \,,
\end{equation}
consisting of a time-direction and a fiber bundle.
Here $t$ is time, $\gamma$ parametrizes a circle fibered over a base space with coordinates $x^i$, metric $h_{ij}$ and U$(1)$ connection $A=A_idx^i$. $R$ is the radius of $\gamma$ and also sets the scale of the fiber bundle.
We assume $\gamma$ has period $2\pi$ and that $\partial_t$ and $\partial_\gamma$ are Killing vectors.

Our focus in the following is on when the U$(1)$ connection $A$ is non-flat $dA \neq 0$. This means the circle fiber is geometrically twisted in the sense that going around in a closed loop in the base space changes the angle $\gamma$ by the amount $\oint A$.%
\footnote{Some of the fiber bundles considered in this paper are topologically non-trivial as well with non-vanishing first Chern class. This means there exists a closed two-cycle $C$ such that $\frac{1}{2\pi} \int_C dA$ is a non-zero integer.}

Having $\gamma-t$ and $x^i$ fixed is a null worldline. Assuming the geometry is part of a target space in string theory, this null worldline can correspond to a BPS state with supersymmetry.
As we shall see, the geometric twist of the circle fiber 
has the consequence that  the BPS decoupling limit associated with the BPS state is not a DLCQ limit, unlike in Section \ref{sec:no_twist}.

As a first example arising from the AdS/CFT correspondence, we consider the Hopf fibration of $S^3$ with base space $\CP^1$ (one-dimensional complex projective space) and connection given by
\begin{equation}
\label{metpot_S3}
\begin{array}{c} \ds
h_{ij} dx^i dx^j = d\Sigma_1^{\, 2} = \frac{1}{4} ( d\theta^2 + \sin^2 \theta \, d\phi^2 )  \,,
\\[4mm] \ds
A = \frac{1}{2}\cos \theta \, d\phi  \,.
\end{array}
\end{equation}
With this, the background \eqref{met_onlyA} is $\R \times S^3$ which - as we shall see below - emerges naturally from AdS$_5\times S^5$. The null worldline corresponds in this case to a $1/4$ BPS state \cite{Mateos:2003de}.
The U$(1)$ connection $A$ in \eqref{metpot_S3} is clearly non-flat and hence the circle fiber is geometrically twisted. It is also topologically non-trivial since  $\int_{\CP^1} dA =- 2\pi$ in line with the $S^3$ topology of the fiber bundle.

Introduce now the energy $E = i \partial_t$
and the momentum $p_\gamma = - i \partial_\gamma$ both measured in units of $(\sqrt{\alpha'}R)^{-1}$.
With a non-zero momentum $p_\gamma=w$, $w$ being a positive integer, one has the bound $E \geq w$ saturated by the BPS state $E=w$.
For $S^3$, $p_\gamma=J_1+J_2$ where $J_1$ and $J_2$ are the angular momenta in the two planes in the embedding $S^3 \subset \R^4$.

The aim in the following is to find a decoupling limit that captures the dynamics of the fluctuations of a fundamental string near the BPS state. 
Therefore, we follow the same strategy as for the untwisted limit in Section \ref{sec:no_twist} by introducing the new coordinates $T$ and $\hat{\gamma}$
\begin{equation}
\label{coords_onlyA}
t = \frac{T}{\epsilon^2} \spa \gamma = \hat{\gamma} + \frac{T}{\epsilon^2} \,,
\end{equation}
which we will keep finite below in the limit $\epsilon\rightarrow 0$. 
The metric is then
\begin{equation}
\label{met_onlyAb}
\frac{ds^2}{\alpha' R^2} =   2 ( d\hat{\gamma} + A) \frac{dT}{\epsilon^2} + (d\hat{\gamma} + A)^2 + h_{ij} dx^i dx^j \,.
\end{equation}
In this frame, the energy associated with $T$ is $E_T = (E - w)/\epsilon^2$ and $\hat{\gamma}$ is periodic with period $2\pi$. Moreover, the null worldline of the BPS state is parametrized by $T$ with $\hat{\gamma}$ and $x^i$ held fixed.

Zooming in on the fluctuations close to BPS means $\epsilon \rightarrow 0$ with $E_T$ held fixed. The question is how $R$ should scale in this limit. To analyze this, we consider a string probe and perform a T-duality
along the circle direction $\hat{\gamma}$. Just as in the DLCQ case, this will clarify the scaling of $R$ and also reveal what type of target-space geometry the surviving modes in the BPS decoupling limit will probe.
From the Buscher rules a T-duality along $\hat{\gamma}$ gives the background
\begin{equation}
\label{Tdualmet_onlyA}
\frac{ds^2}{\alpha' } = - \frac{R^2}{\epsilon^4} dT^2 + \frac{dv^2}{R^2}  + R^2 h_{ij} dx^i dx^j \,,
\end{equation}
\begin{equation}
\label{Bfield_onlyA}
\frac{B}{\alpha'} = - \left( \frac{dT}{\epsilon^2} + A \right)\wedge dv  \,,
\end{equation}
where the T-dual circle direction $v$ is periodic with period $2\pi$. Notice the T-duality has untwisted the fiber bundle geometry, transferring the connection $A$ to the Kalb-Ramond field.

Consider now the Nambu-Goto action 
\begin{equation}
\label{NGaction}
S = - \frac{1}{2\pi \alpha'} \int d^2\sigma \left( \sqrt{-\det(g_{\alpha\beta})} + \frac{1}{2} \epsilon^{\alpha\beta} B_{\alpha\beta} \right) \,,
\end{equation}
with worldvolume coordinates $\sigma^0=\tau$ and $\sigma^1=\sigma$ and pullback given as $g_{\alpha\beta} = \partial_\alpha x^\mu \partial_\beta x^\nu g_{\mu\nu}$ with $\alpha,\beta=0,1$ and $B_{\alpha\beta}$ is defined similarly.
We choose the gauge $T = \tau$ and $v = w \,\sigma$ in accordance with having winding number $w$ along $v$. 

In the BPS decoupling limit $\epsilon\rightarrow 0$ the square root term in \eqref{NGaction} becomes
\begin{equation}
\frac{\alpha' w}{\epsilon^2} \sqrt{ \left(1-\epsilon^4 h_{00}\right) \left(1+\frac{R^4}{w^2} h_{11} \right) + \frac{\epsilon^4 R^4}{w^2} h_{01}^2 } \,,
\end{equation}
where we used the pullback $h_{\alpha\beta}= \partial_\alpha x^i \partial_\beta x^j h_{ij}$.
We need $R\rightarrow 0$ for the $\epsilon^{-2}$ divergence in $B$ to be canceled. Moreover, to avoid further divergences $R$ should go to zero at least as $\sqrt{\epsilon}$. Demanding both that the limit be finite and that the resulting action should couple to $h_{ij}$ therefore requires that $R$ scales as $\sqrt{\epsilon}$.
Thus, we have derived the scaling
\begin{equation}
\label{twisted_scaling}
R = \sqrt{\epsilon}
\end{equation}
up to a multiplicative constant that will not affect the physics.
With this, one obtains from \eqref{NGaction} the finite action
\begin{equation}
\label{finite_S_onlyA}
S = \frac{w}{2\pi} \int d^2\sigma \left( 
A_i \dot{x}^i  - \frac{1}{2w^2} h_{ij} x'{}^i x'{}^j 
 \right)   \,,
 \end{equation}
where dot and prime are the worldvolume derivatives $\partial_0$ and  $\partial_1$, respectively. One has in addition a momentum constraint $\int_0^{2\pi} d\sigma A_i x'{}^i = 0$ arising from the fact that no winding along the T-dual direction $\gamma$ implies no momentum along $v$.

We call the limit with scaling \eqref{twisted_scaling} a
 {\sl twisted BPS decoupling limit} in that it zooms in on BPS and near-BPS excitations of a string for the BPS momentum state $p_\gamma=w$ along a circle fiber which is geometrically twisted on the base space. 
 
This limit is different from a DLCQ limit. 
Indeed, the radius $R$ of $\hat\gamma$ scales differently with $\epsilon$ than in the DLCQ limit of Section \ref{sec:no_twist}. 
Also, a DLCQ limit would require $\hat{\gamma}$ to become a null-coordinate in the $R=\sqrt{\epsilon} \rightarrow 0$. However, considering the background \eqref{met_onlyAb} we notice that most of the metric components go to zero as $\epsilon$, except for  $g_{\hat{\gamma} T}$ and $g_{\phi T}$. This should be interpreted as the effective string tension going to zero as $\epsilon$. 
Finally, another important distinction from the DLCQ limit 
is that while the ISO$(1,1)$ boost symmetry $\gamma \partial_t + t\partial_\gamma$
is a Killing vector field of \eqref{circle1} the 
coupling of $\gamma$ to the connection $A$ in \eqref{met_onlyA} breaks this symmetry.

The sigma-model \eqref{finite_S_onlyA} was previously derived from a two-step limiting procedure in \cite{Harmark:2017rpg,Harmark:2018cdl,Harmark:2020vll,Bidussi:2023rfs}. We will compare this procedure to the twisted BPS decoupling limit at the end of this section.

For the $\R \times S^3$ background with Hopf fibration \eqref{metpot_S3}, the finite sigma-model \eqref{finite_S_onlyA} in the twisted BPS decoupling limit becomes the Landau-Lifshitz sigma-model
\begin{equation}
\label{LLmodel}
S= \frac{w}{4\pi} \int d^2\sigma \left( 
\cos \theta \, \dot{\phi}  - \frac{1}{4w^2} \Big( \theta'{}^2 + \sin^2 \theta \, \phi'{}^2 \Big) 
 \right) \,.
 \end{equation}
How does this arise from the full AdS$_5\times S^5$ background?
To see this, start with the metric of AdS$_5\times S^5$ in the global patch, here written as
\begin{eqnarray}
\label{metric_fullsu2}
ds^2 &=& \alpha' R^2 \left[ - \cosh^2 \! \rho \,dt^2 + d\rho^2 + \sinh^2 \! \rho \, d\widetilde{\Omega}_3^2 + d\xi^2 \right. \nn \\ && \left.  + \cos^2 \! \xi \Big( d\Sigma_1^{\, 2} + (d\gamma + A)^2 \Big) +\sin^2 \! \xi \,d\chi^2 \right]  \,,
\end{eqnarray}
with $d\widetilde{\Omega}_3^{\, 2}$ the unit three-sphere metric.
One has a $1/4$ BPS state associated with the null worldline defined by $\gamma-t$ fixed, keeping $\rho=0$, $\xi=0$ and with fixed angles $(\theta,\phi)$ on $\CP^1$.  
Using the coordinate transformation \eqref{coords_onlyA} and performing a T-duality along $\hat{\gamma}$ we get the type IIA string theory background
\begin{eqnarray}
\frac{ds^2}{\alpha'} &=& \frac{dv^2}{R^2\cos^2 \! \xi}+ R^2 \left[ - \cosh^2 \! \rho  \frac{dT^2}{\epsilon^4}
+ \cos^2 \xi \, d\Sigma_1^{\, 2}  
   \right.
\nn \\ 
&& \left.
+ d\rho^2+ \sinh^2 \! \rho \, d\widetilde{\Omega}_3^2 +d \xi^2 + \sin^2 \! \xi \, d\chi^2
\right] \,,
\end{eqnarray}
with Kalb-Ramond field \eqref{Bfield_onlyA} and dilaton $g_s e^\phi = \frac{R^3}{4\pi N \cos \xi}$.

Considering the Nambu-Goto action \eqref{NGaction} in the gauge $T=\tau$ and $v = w\, \sigma$ 
we notice that in the limit $R=\sqrt{\epsilon} \rightarrow 0$ the divergent part of the $B$-field contributes with $- w/\epsilon^2$ while the divergent part of the square root term is $w \cosh \rho /( \cos \xi \, \epsilon^2)$. Hence, 
cancellation of the divergence requires
 $\rho=\xi=0$. This means the surviving perturbative string degrees of freedom live on the $\mathbb{R} \times S^3$ subspace defined by $\rho=\xi=0$, since the divergence gives an infinitely steep potential that confines the modes to this subspace \cite{Harmark:2008gm,Harmark:2017rpg,Harmark:2018cdl}.
Below in Section \ref{sec:Dbranes} we shall see that D-branes, on the other hand, can probe $\rho > 0$ in the same twisted BPS decoupling limit.

Remarkably, one can obtain the same sigma-model \eqref{LLmodel} starting from $\CN=4$ supersymmetric Yang-Mills (SYM) with gauge group SU$(N)$ in the planar limit $N=\infty$ (corresponding to tree-level string theory $g_s=0$). This is done by translating the twisted BPS decoupling limit $R^2=\epsilon \rightarrow 0$ to a limit of  $\CN=4$ SYM, known as the SU$(2)$ Spin Matrix theory (SMT) limit \cite{Harmark:2014mpa}.
Building on the seminal work of  \cite{Minahan:2002ve}, one finds that planar $\CN=4$ SYM
in the SU$(2)$ limit reduces to a Heisenberg XXX$_{1/2}$ spin chain of length $w$ \cite{Harmark:2008gm}. 
Considering a long-wavelength limit with $w \gg 1$ this spin chain can be described as the sigma-model \eqref{LLmodel} known as the Landau-Lifshitz sigma-model \cite{Fradkin1991,Kruczenski:2003gt}. Thus, one has a direct bridge between the gauge theory side and the string theory side of the holographic correspondence in this limit. This works because the limit zooms in on the dynamics close to a BPS state, avoiding problems with zero-mode fluctuations, because the AdS$_5\times S^5$ background is protected against $\alpha'$ corrections and because one compares the sigma-models in a semi-classical regime \cite{Harmark:2008gm}.%
\footnote{A related way to connect emerging spin chains from integrable sectors of planar $\CN=4$ SYM with limits of strings on AdS$_5\times S^5$ was pointed out in  \cite{Kruczenski:2003gt} and expanded on in \cite{Kruczenski:2004kw,Hernandez:2004uw,Stefanski:2004cw,Bellucci:2004qr}, in particular obtaining the Landau-Lifshitz model. However, the limit of \cite{Kruczenski:2003gt} 
on the string theory side sends $J$ to infinity while keeping $R^4/J$ fixed. This means one keeps higher orders in $R^4/J$ and thus one is not taking a limit that zooms in toward the $1/4$-BPS state $E=J$. Therefore, it does not correspond to a BPS decoupling limit.}

Previously, the string dual of the SU$(2)$ SMT limit has been considered in \cite{Harmark:2017rpg,Harmark:2018cdl,Harmark:2020vll,Bidussi:2023rfs}.
However, in these works the sigma-model \eqref{LLmodel}  arose from a two-step limiting procedure. In \cite{Harmark:2017rpg,Harmark:2018cdl,Harmark:2020vll} a DLCQ-type procedure was first performed with $t = x^0 - \frac{1}{2} u$ and $\gamma = x^0 + \frac{1}{2} u$ where $u$ is a null-coordinate. Subsequently the SMT limit was realized by sending $\epsilon \rightarrow 0$ with $x^0 = \tilde{x}^0 /\epsilon^2$. The combination of these two limits amounts to the same resulting sigma-model as the above twisted BPS decoupling limit since comparing with \eqref{coords_onlyA} we find that the coordinates are connected as $T= \tilde{x}^0 - \frac{1}{2} \epsilon^2 u$ and $\hat{\gamma}=u$ which means $E_T = E_{\tilde{x}^0}$ and $P_u = J + \frac{1}{2}\epsilon^2 E_T$. However, the initial DLCQ-like procedure of \cite{Harmark:2017rpg,Harmark:2018cdl,Harmark:2020vll,Bidussi:2023rfs} is not an actual BPS decoupling limit since $u$ is not a compact coordinate, and since $P_u$ is not the correct momentum to keep fixed.
Therefore, this DLCQ-like procedure is not by itself one that zooms in toward a BPS state. It is only the combination of the two limits, which we have shown above corresponds to a twisted BPS decoupling limit, that zooms in on a BPS state.

\section{General twisted limit and target space geometry}
\label{sec:genlimit}

In this section we present a general class of twisted BPS decoupling limits. 
Subsequently, we discuss the emergent target space geometry that string fluctuations probe in such limits. 

\subsection{General twisted limit}

In Section \ref{sec:twistedlimit} we chose as starting point the geometry \eqref{met_onlyA} which is a product of a flat time direction and a fiber bundle. 
We now consider a more general starting point with the following class of $(d+2)$-dimensional backgrounds
\begin{equation}
\label{gen_metric_hat}
\frac{ds^2}{\alpha'} = - (dT - \tilde{A})^2 + ( d\hat{\gamma} + dT+A)^2 + h_{ij} dx^i dx^j  \,,
\end{equation}
where $\partial_T$ and $\partial_{\hat{\gamma}}$ are assumed to be Killing vector fields. 
Here we allow for a fiber bundle structure both with respect to a circle fiber $\hat{\gamma}$ and a time-direction fiber $T$. For this reason,  in addition to the connection $A=A_i dx^i$ for the $\hat{\gamma}$ fiber, we also introduce a connection 
$\tilde{A}=\tilde{A}_i dx^i$ for the $T$ fiber. Therefore, we take the metric $h_{ij}$, with coordinates $i,j=1,...,d$, to describe a combined base space geometry in a fiber bundle for which both $\hat{\gamma}$ and $T$ are fibers. 
The metric \eqref{gen_metric_hat} is seen to generalize \eqref{met_onlyAb} to include $\tilde{A}$. In Section \ref{sec:examples} we shall see examples for the AdS$_5\times S^5$ background that includes geometries \eqref{gen_metric_hat} with non-flat $\tilde{A}$ connections.

We assume that the null worldline obtained by holding $x^i$ and $\hat{\gamma}$ fixed corresponds to a BPS state. Here $T$ parametrizes the worldline.
Moreover, the momentum $w = -i \partial_{\hat{\gamma}}$ is assumed positive.

We have in general $d \leq 8$. When $d < 8$ it means we are considering a subspace of ten-dimensional type II supergravity on which $B$ is critical, similarly to what happened in the example of Section \ref{sec:twistedlimit} in which a string on AdS$_5\times S^5$ became confined to $\mathbb{R} \times S^3$.

Performing now a T-duality along $\hat{\gamma}$ we obtain the background
\begin{equation}
\label{gen_metric_B}
\begin{array}{c} \ds
\frac{ds^2}{\alpha'} = - (dT - \tilde{A})^2  + h_{ij} dx^i dx^j    +  dv^2 \,,
\\[4mm] \ds
\frac{B}{\alpha'} = -  (dT + A ) \wedge dv \,.
\end{array}
\end{equation}
In this T-dual frame the BPS state corresponds to a winding mode with winding number $w >0$ along $v$.

The twisted BPS decoupling limit of \eqref{gen_metric_B}  is now defined as $\epsilon \rightarrow 0$ after performing the substitution
\begin{equation}
\label{limit_sub}
\begin{array}{c} \ds
T \rightarrow \epsilon^{-\frac{3}{2}} T \spa v \rightarrow \epsilon^{-\frac{1}{2}} v \,,
\\[2mm] \ds
h_{ij} \rightarrow \epsilon \, h_{ij} \spa A \rightarrow \sqrt{\epsilon} \, A \spa \tilde{A} \rightarrow \sqrt{\epsilon} \, \tilde{A}  \,,
\end{array}
\end{equation}
with $i,j=1,2,...,d$. 
In this limit, the Nambu-Goto action \eqref{NGaction} becomes
\begin{equation}
\label{galilean_action_fixed}
\begin{array}{c} \ds
S = \frac{w}{2\pi} \int d^2 \sigma 
 \left( m_i \dot{x}^i - 
\CH \right)  \,,
\\[4mm] \ds
 \CH (x,x') = \frac{1}{2w^2} h_{ij} x'{}^i x'{}^j \,,
\end{array}
\end{equation}
where we have defined the one-form
\begin{equation}
m = A + \tilde{A} \,.
\end{equation}
We have chosen the gauge
\begin{equation}
T = \tau \spa v= w\,  \sigma \,,
\end{equation}
with respect to the worldvolume coordinates $\tau$ and $\sigma$ where $w > 0$ is the winding number. 
As this gauge implies $\dot{v}=0$ we have the constraint
\begin{equation}
\label{momentum_con}
\int_0^{2\pi} d\sigma\,  m_i x'{}^i = 0 \,,
\end{equation}
expressing the total momentum along $v$ is zero.

\subsection{Symplectic nature of target space}
\label{sec:symplectic}

The action \eqref{galilean_action_fixed} has features similar to a phase-space action for a particle. Indeed, applying the Euler-Lagrange equations to \eqref{galilean_action_fixed} we get 
\begin{equation}
\label{time_evol}
\omega_{ij} \dot{x}^j = \frac{\partial \CH}{\partial x^i} - \left( \frac{\partial \CH}{\partial x'{}^i} \right)' \,,
\end{equation}
where we defined the two-form
\begin{equation}
\omega = dm \,.
\end{equation}
Assume now $\omega$ to be non-degenerate, or, equivalently, $\omega_{ij}$ to be invertible.
Then  Eq.~\eqref{time_evol}
 suggests we should view $x^i$ as coordinates on phase space, with $\omega$ being the symplectic form. 
 This is seen from the fact that it 
links the time evolution of $x^i$ to the Hamiltonian $H=\int_0^{2\pi} d\sigma \CH$.
Thus, in the twisted BPS decoupling limit the target space geometry becomes a symplectic geometry \cite{Harmark:2008gm,Harmark:2020vll,Bidussi:2023rfs}.
In particular, the one-form $m$ is the symplectic potential for $\omega$. 
From  \eqref{time_evol} one finds  the Poisson bracket with respect to $H$ as
\begin{equation}
\{ x^i(\tau,\sigma) , x^j (\tau,\sigma' ) \} = \Delta^{ij} (\sigma) \delta (\sigma-\sigma') \,,
\end{equation}
where we defined
\begin{equation}
\int_0^{2\pi} d\hat{\sigma} \Delta^{ij} (\sigma-\hat{\sigma}) \omega_{jk}  (\hat{\sigma}-\sigma') = \delta^i_k \delta (\sigma-\sigma') \,.
\end{equation}
The phase-space formulation of the action \eqref{galilean_action_fixed}
 has been further explored for a solvable class of space-times in 
\cite{Bidussi:2023rfs}.

For the twisted limit of Section \ref{sec:twistedlimit} giving the sigma-model \eqref{LLmodel} the target space geometry has
$m= \frac{1}{2} \cos \theta \, d\phi$ and $h_{\mu\nu}dx^\mu dx^\nu=d\Sigma_1^{\, 2}$. This is in line with the symplectic nature of the target space in the limit, as $\CP^1$ is a K\" ahler geometry with $h_{ij}$ being the Fubini-Study metric and $\omega$ the  K\" ahler form. We shall see in Section \ref{sec:examples} all twisted BPS decoupling limits on AdS$_5\times S^5$ have a symplectic K\" ahler structure.

\subsection{Non-Lorentzian geometry}

The above formulation of the string dynamics in a twisted BPS decoupling limit is coordinate dependent. It is interesting to consider a covariant formulation to fully exhibit the geometry that a string will probe in the limit. To this end, consider the covariant version of the background \eqref{gen_metric_B}
\begin{equation}
\label{Cov_metric_B}
\begin{array}{c} \ds
\frac{ds^2}{\alpha'} = - (\tau^0 - \tilde{A})^2+  (\tau^1)^2  + h_{\mu\nu} dx^\mu dx^\nu     \,,
\\[4mm] \ds
\frac{B}{\alpha'} = -  (\tau^0 + A ) \wedge \tau^1 \,,
\end{array}
\end{equation}
with $d+2$ coordinates $x^\mu$ and indices $\mu,\nu = 0,1,...,d+1$. We assume $d\tau^a = 0$.
We write $h_{\mu\nu} = \delta_{ij} e^i_\mu e^j_\nu$ with the transverse vielbeins $e^i_\mu$ with $i,j=1,2,...,d$. The twisted BPS decoupling limit is now $\epsilon\rightarrow 0$ after the substitution
\begin{equation}
\label{limit_sub2}
\begin{array}{c} \ds
\tau^0 \rightarrow \epsilon^{-\frac{3}{2}} \tau^0  \spa \tau^1 \rightarrow \epsilon^{-\frac{1}{2}}  \tau^1 \,,
\\[2mm] \ds
e^i \rightarrow \sqrt{\epsilon} \, e^i \spa A \rightarrow \sqrt{\epsilon} \, A \spa \tilde{A} \rightarrow \sqrt{\epsilon} \, \tilde{A}  \,,
\end{array}
\end{equation}
with $i=1,2,3,...,d$. This provides a coordinate-independent version of the limit.

To exhibit the $\epsilon \rightarrow 0$ limit of the Nambu-Goto action \eqref{NGaction}, we introduce
\begin{equation}
\tau^a_\alpha = \partial_\alpha x^\mu \tau^a_\mu
 \spa \tau^\alpha_ a = - \frac{\epsilon^{\alpha\beta} \epsilon_{ab} \tau^b_\beta}{\sqrt{-\tau}}
 \spa 
\sqrt{-\tau} = \epsilon^{\alpha\beta} \tau^0_\alpha \tau^1_\beta \,,
\end{equation}
with $a,b=0,1$ and $-\epsilon_{01} = \epsilon^{01} = 1$ so that $\tau_\alpha^a \tau_a^\beta = \delta_\alpha^\beta$ and $\tau_\alpha^a \tau_b^\alpha = \delta_b^a$.
The winding mode along $v$ implies $\tau^a_\alpha$ is invertible.
With this, the square root term in \eqref{NGaction} gives
\begin{equation}
\sqrt{-\tau} \left(
\frac{1}{\epsilon^2} -  \tau_0^\alpha \tilde{A}_\alpha + \frac{1}{2} \tau_1^\alpha \tau_1^\beta  h_{\alpha\beta} + \CO(\epsilon^2) \right)
 \,.
 \end{equation}
Notice the divergent piece cancels the one from the $B$-field. The finite action that results from the limit is \cite{Harmark:2017rpg,Harmark:2018cdl,Bidussi:2023rfs}%
\footnote{In  \cite{Bidussi:2023rfs} the action \eqref{galilean_action} is derived, though with the important difference that we impose additional conditions of exactness of $\tau^a$ and that $b$ in our case is given by Eq.~\eqref{b_def}. Ref.~\cite{Bidussi:2023rfs} employs a two-step limiting procedure, consisting first of a $c\rightarrow \infty$ limit corresponding to the NRST limit of \cite{Bidussi:2021ujm} and subsequently a longitudinal speed of light limit $\tilde{c}\rightarrow \infty$ resulting in the action \eqref{galilean_action}. Comparing to the one-step limit of this work we identify $c=\epsilon^{-2}$ and $\tilde{c}=\epsilon^{-1}$. 
If one fixes $\tau^1 = dv$ the action \eqref{galilean_action} reduces to the one derived in \cite{Harmark:2017rpg,Harmark:2018cdl}.
Finally, note the intriguing observation of \cite{Gomis:2023eav} that one can obtain an action of the same form from a double-DLCQ limit (though again without the constraints $d\tau^a=0$ and Eq.~\eqref{b_def}, which is part of the definition of U$(1)$-Galilean geometry).}
\begin{equation}
\label{galilean_action}
S = - \frac{1}{4\pi} \int d^2\sigma 
\left( \epsilon^{\alpha\beta} b_{\alpha\beta}  +  \sqrt{-\tau}\,\tau_1^\alpha \tau_1^\beta  h_{\alpha\beta}\right) \,,
\end{equation}
with $b_{\alpha\beta}$ being the pullback of the two-form
\begin{equation}
\label{b_def}
b = \tau^1 \wedge m   =  \tau^1 \wedge ( A + \tilde{A} )  \,.
\end{equation}
The winding of the string is fixed by
\begin{equation}
w = \frac{1}{2\pi} \int_0^{2\pi} d\sigma \, \partial_\sigma x^\mu \tau^1_\mu  \,.
\end{equation}

Using the analysis of \cite{Bidussi:2021ujm,Bidussi:2023rfs} we find that the action \eqref{galilean_action} has the local infinitesimal Galilean boost symmetry
\begin{equation}
\label{boost}
\delta \tau^a = 0 \spa \delta m = 0 \spa
\delta h_{\mu\nu} = \lambda_{i} ( e^i _\mu \tau^0 _\nu +e^i _\nu \tau^0 _\mu ) \,,
\end{equation}
with local boost parameter $\lambda_i$.
In addition one has diffeomorphisms for the $x^\mu$ coordinates and U$(1)$ gauge transformations of the one-form $m$. 
The non-Lorentzian geometry that these local symmetries define is known as U$(1)$-Galilean geometry \cite{Harmark:2017rpg,Harmark:2018cdl}.
Additional local symmetries discussed in \cite{Bidussi:2021ujm,Bidussi:2023rfs} are broken due to demanding the specific form for $b$ in Eq.~\eqref{b_def} and $d\tau^a=0$.
This defines the target space geometry that the near-BPS modes of the string are probing in a twisted BPS decoupling limit.

\section{Twisted limits of AdS$_5\times S^5$}
\label{sec:examples} 

In this section we consider all of the twisted BPS decoupling limits of the AdS$_5\times S^5$ background of type IIB string theory, of which we saw an example in Section \ref{sec:twistedlimit}.
Write the metric of AdS$_5\times S^5$ in the global patch as
\begin{equation}
\label{gen_adss5}
\frac{ds^2}{\alpha' R^2} = - \cosh^2 \! \rho \, dt^2 + d\rho^2 + \sinh^2 \! \rho \, d\widetilde{\Omega}_3^{\, 2}
 + d\Omega_5^{\, 2} \,.
\end{equation}
This background has global isometry group PSU$(2,2|4)$. The Cartan charges are the energy $E=-i\partial_t$, the angular momenta $S_1,S_2$ of $d\widetilde{\Omega}_3^{\, 2}$  and the angular momenta $J_1,J_2,J_3$ of $d\Omega_5^{\, 2}$.
The BPS states of PSU$(2,2|4)$ obey
\begin{equation}
\label{BPS_form}
E = \sum_{i=1}^{p} S_i + \sum_{j=1}^{q+1} J_j \,,
\end{equation}
with $p,q\geq 0$ and either $p=q=0$ ($1/2$ BPS), $p+q=1$ ($1/4$ BPS), $p+q=2$ ($1/8$ BPS) or $p=q=2$ ($1/16$ BPS).

Each BPS state \eqref{BPS_form} corresponds to a null worldline of AdS$_5\times S^5$. Given $p$ and $q$, the BPS null worldline is contained in an AdS$_{2p+1} \times S^{2q+1}$ subspace of AdS$_5\times S^5$ (identifying AdS$_1$ with the flat time direction $\mathbb{R}$). This is discussed in detail in Appendix \ref{app:null_world}. As a result, one can write the metric of the subspace as 
\begin{eqnarray}
\label{subspace_pq}
\frac{ds^2}{\alpha'  R^2} &=& - \left( \frac{dT}{\epsilon^2}     - \tilde{A}^{(p)} \right)^2  + \left(\frac{dT}{\epsilon^2} + d\hat{\gamma} + A^{(q)} \right)^2  
\nn \\[2mm] &&
+ dH_p^{\, 2}+  d\Sigma_q^{\, 2} \,,
\end{eqnarray}
where the BPS state corresponds to the null worldline parametrized by $T$ with all other coordinates held fixed. 
Here $dH_p^{\, 2}$ is the metric of $p$-dimensional complex hyperbolic space  $\mCH^p$ 
with K\" ahler one-form potential $\tilde{A}^{(p)}$.%
\footnote{$p$-dimensional complex hyperbolic space $\mCH^p$ is the space of negative complex lines in $\bbC^{p,1}$. Here $\bbC^{p,1}$ is $\bbC^{p+1}$ with a norm $|z|^2 = |z_1|^2+\cdots + |z_p|^2-|z_{p+1}|^2$ of signature $(1,...,1,-1)$.} 
Moreover, $d\Sigma_q^{\, 2}$ is the Fubini-Study metric of $q$-dimensional complex projective space with 
K\" ahler one-form potential $A^{(q)}$. See Appendix \ref{app:null_world} for explicit expressions.
For $p=0$ and $q=1$ the metric \eqref{subspace_pq} reduces to the special case given by Eqs.~\eqref{metpot_S3} and  \eqref{met_onlyAb} describing $\mathbb{R} \times S^3$ (with $A=A^{(1)}$). 

The energy associated with $T$ in \eqref{subspace_pq} is
the rescaled energy above the BPS state
$E_T = ( E -  \sum_{i=1}^p S_i -\sum_{j=1}^{q+1} J_j)/\epsilon^2$. The $\hat{\gamma}$ direction is periodic with period $2\pi$ and we have $-i\partial_{\hat{\gamma}} = \sum_{i=1}^{q+1} J_i$.
Note that for $p=0$ we identify $\tilde{A}^{(0)}=0$ and $dH_0^{\, 2}=0$ while for $q=0$ we identify $A^{(0)}=0$ and $d\Sigma_0^{\, 2}=0$.

Following Section \ref{sec:genlimit}, we now proceed to T-dualize along the $\hat{\gamma}$ direction. This keeps fixed 
\begin{equation}
\label{def_w}
w=\sum_{i=1}^{q+1} J_i \,,
\end{equation}
as the winding number of the T-dual direction $v$. 
Choosing this direction for T-duality, also for $p=1,2$, is in line with the connection to the spin chain picture of planar $\CN=4$ SYM in which the $S_1$ and $S_2$ quantum numbers count excitations on the sites of the chain rather than contributing to the length of the chain.
We then obtain the T-dual background 
\begin{equation}
\label{met_Tdual_pq}
\frac{ds^2}{\alpha' } = R^2 \left[ - \left( \frac{dT}{\epsilon^2} - \tilde{A}^{(p)}\right)^2+   dH_p^{\, 2} +  d\Sigma_q^{\, 2}\right]
 + \frac{dv^2}{R^2} \,,
\end{equation}
with Kalb-Ramond field
\begin{equation}
\label{B_Tdual_pq}
\frac{B}{\alpha'} =  - \left( \frac{dT}{\epsilon^2} + A^{(q)} \right) \wedge dv \,.
\end{equation}
Here the metric \eqref{met_Tdual_pq} describes AdS$_{2p+1}\times \CP^q \times S^1$ geometry. In case $p=1,2$  AdS$_{2p+1}$ is described as a fiber bundle with base space 
$\mCH^p$ and the time-direction $T$ parameterizing a geometrically twisted fiber with connection $\tilde{A}^{(p)}$. 
Instead, the T-duality has untwisted the $S^{2q+1}$ geometry to $\CP^q \times S^1$, with the connection $A^{(q)}$ instead being part of the Kalb-Ramond field \eqref{B_Tdual_pq}.

The twisted BPS decoupling limit associated with the BPS state \eqref{BPS_form} is taken as $R^2=\epsilon\rightarrow 0$, following the general prescription
of Eqs.~\eqref{gen_metric_B}-\eqref{limit_sub}.
Taking this limit of the 
 Nambu-Goto action \eqref{NGaction} on the background \eqref{met_Tdual_pq}-\eqref{B_Tdual_pq} 
 zooms in on stringy fluctuations very close to 
 the BPS state. By taking the limit we see that the near-BPS fluctuations are
 described by the non-relativistic sigma-model action \eqref{galilean_action} 
 in the U$(1)$-Galilean background 
\begin{equation}
\label{bg_su_pq}
\begin{array}{c}
\tau^0 = dT \spa  \tau^1 = dv  \,,
\\[2mm]
m = \tilde{A}^{(p)} + A^{(q)} \spa h_{\mu\nu}dx^\mu dx^\nu = dH_p^{\, 2} + d\Sigma_q^{\, 2} \,.
\end{array}
\end{equation}
This is a $\mCH^p \times \CP^q$ K\" ahler geometry with metric $h_{\mu\nu}$ and symplectic form $\omega = dm$ in line with the interpretation of the U$(1)$-Galilean geometry as symplectic. This background has isometry group SU$(1,p) \times \mbox{SU}(q+1)$.
For $p=0$ it reduces to a $\CP^q$ geometry with isometry group SU$(q+1)$. 

Turning on the transverse directions to the AdS$_{2p+1} \times S^{2q+1}$ subspace of AdS$_5\times S^5$ moves one away from the region in which the divergence of the square root term cancels with the $B$-field term in the Nambu-Goto action \eqref{NGaction}. In the limit $R^2=\epsilon \rightarrow 0$ this works as an infinitely steep confining potential, restricting the dynamics to be infinitesimally close to the null worldline in the AdS$_{2p+1} \times S^{2q+1}$ subspace. 

The U$(1)$-Galilean backgrounds \eqref{bg_su_pq} are equivalent to those found by the above-mentioned two-step procedure in \cite{Harmark:2008gm,Harmark:2017rpg,Harmark:2020vll}.%
\footnote{Note that the identification of the general subspace structure as well as the complex hyperbolic space geometric structure for $p=1,2$ is new.}
As explained above, the new procedure of this work provides a direct route to the near-BPS dynamics, avoiding the intermediate non-BPS DLCQ-like limit.

{\bf Gauge theory dual of twisted limits:} 
The twisted BPS decoupling limits of AdS$_5\times S^5$ exhibited above are dual to Spin Matrix theory (SMT) limits of $\CN=4$ SYM on $\mathbb{R}\times S^3$ \cite{Harmark:2014mpa,Harmark:2019zkn,Baiguera:2020jgy,Baiguera:2020mgk,Baiguera:2021hky,Baiguera:2022pll}.
Indeed, given a BPS state \eqref{BPS_form} 
 the Spin Matrix theory limit is 
\begin{equation}
\label{SMT_limit}
\lambda \rightarrow 0\ \ \mbox{with} \ \ \frac{E-\sum_{i=1}^p S_i - \sum_{j=1}^{q+1} J_j}{\lambda} = \mbox{fixed} \,.
\end{equation}
Here $E$ is the energy, $S_i$ are angular momenta and $J_j$ are R-charges of $\CN=4$ SYM on $\mathbb{R}\times S^3$ with gauge group SU$(N)$. $N$ is held fixed in the limit.
Since $\lambda$ is translated to $R^4=\epsilon^2$ on the string theory side, the limit \eqref{SMT_limit} is equivalent to the corresponding twisted BPS decoupling limit with fixed rescaled energy $E_T$.

As a consistency check, we note that the isometry group SU$(1,p) \times \mbox{SU}(q+1)$ 
of the background \eqref{bg_su_pq}
can be matched to the bosonic part of the global symmetry of SMT \cite{Harmark:2014mpa,Harmark:2019zkn,Baiguera:2020jgy,Baiguera:2020mgk,Baiguera:2021hky,Baiguera:2022pll}. For the cases that include fermions, the global symmetry group is enhanced to a super Lie group.

Above we considered the twisted BPS decoupling limit of AdS$_5\times S^5$ for a single string probe, resulting in the sigma-model action \eqref{galilean_action} on the U$(1)$-Galilean background \eqref{bg_su_pq}. 
This matches perfectly what one finds on the gauge theory side \cite{Harmark:2008gm,Harmark:2017rpg,Harmark:2018cdl,Harmark:2020vll}, as we now review, extending the match mentioned in Section \ref{sec:twistedlimit}. 

Since we are taking the twisted limit of a non-interacting single string, we should consider the planar regime of $\CN=4$ SYM. In this regime, single-trace operators are known to have a spin chain description \cite{Minahan:2010js}. 
In the SMT limits, the dynamics further simplifies to nearest-neighbor spin chains \cite{Harmark:2014mpa}.
Moreover, we are considering a semi-classical limit for which $w \gg 1$. As already reviewed in Section \ref{sec:twistedlimit}, this means one can use a long-wavelength description of the spin chain in terms of sigma-models \cite{Fradkin1991,Kruczenski:2003gt}. 
These sigma-models have been found from the spin chain side for $(p,q)=(0,1)$ \cite{Kruczenski:2003gt,Harmark:2008gm}, $(p,q)=(0,2)$ \cite{Hernandez:2004uw,Stefanski:2004cw}, 
$(p,q)=(1,0)$ \cite{Bellucci:2004qr} and $(p,q)=(1,1)$ \cite{Bellucci:2006bv}. The bosonic parts of these sigma-models perfectly match \eqref{galilean_action} in the background \eqref{bg_su_pq} \cite{Harmark:2008gm,Harmark:2017rpg,Harmark:2018cdl,Harmark:2020vll}. It would be interesting to match the fermionic content as well, which would include the matching of global symmetries.

We would like to stress that the matching of sigma-models arising from the string and gauge theory sides of the AdS/CFT correspondence is highly non-trivial. 
On the gauge theory side, the SMT limit combined with focusing on the emergent sigma-model dynamics, corresponds to the regime \cite{Harmark:2008gm,Harmark:2014mpa}
\begin{equation}
\label{weak_thooft}
E - w \ll \lambda \ll 1 \spa w\gg 1 \spa N \gg 1 \,.
\end{equation}
Thus, we are manifestly within a regime of weak 't Hooft coupling.
On the string theory side, on the other hand, we take a limit that involves the string tension going to zero. One could worry if quantum string effects would spoil the classical result that we found for the sigma-model. However, as argued in \cite{Harmark:2008gm}, this is not the case. First of all, the sigma-model can be kept in the semi-classical regime with $w \gg 1$. Secondly, we are close to a BPS state, which keeps the zero mode fluctuations under control. Finally, the AdS$_5\times S^5$ background is protected against $\alpha'$ corrections. These arguments explain why the non-trivial match is possible. 

Finally, let us consider in detail the two extreme cases $(p,q)=(0,0)$ and $(p,q)=(2,2)$.

{\bf The degenerate case $(p,q)=(0,0)$:}
This case is a degenerate version of the BPS decoupling limit, as the geometric twist is absent. 
The reason for this is the absence of a transverse space. Indeed $d=0$ means it has shrunk to a point. 
This reflects the fact that $E - J_1 \rightarrow 0$ zooms in on $1/2$ BPS states that all obey $E=J_1$. This is due to the fact that any state with $E - J_1>0$ is gapped, as revealed by the confining potential that fixes the string to the point  in transverse space corresponding to $\rho=0$ and $\xi = 0$.
Thus, there are no fluctuations with $E - J_1>0$  surviving in this limit, and this is in line with the fact that the Spin Matrix theory limit of $\CN=4$ SYM in this case gives the free theory of $1/2$ BPS states studied in \cite{Berenstein:2004kk,Harmark:2014mpa}.

{\bf The maximal case $(p,q)=(2,2)$:}
This case is the only one in which no directions are confined. Hence it has $d=8$, corresponding to the original number of transverse directions. 
In this case the background \eqref{met_Tdual_pq}-\eqref{B_Tdual_pq}  with $p=q=2$
 describes AdS$_5\times \CP^2 \times S^1$ but with a Kalb-Ramond field that includes a geometric twist of the $S^1$. This is a background of type IIA string theory, with string coupling  $g_s = R^3/(4\pi N)$
and Ramond-Ramond flux
\begin{equation}
F_{(4)} =  \alpha'{}^{\frac{3}{2}} R^3
\cos \xi \, \sin^3 \xi \, \sin \theta \,
  d \xi \, d\eta \, d\theta \, d\phi \,.
\end{equation}
The twisted BPS decoupling limit zooms in on fluctuations near a $1/16$ BPS state with $E=S_1+S_2+J_1+J_2+J_3$.

On the gauge theory side of the AdS/CFT correspondence, the twisted BPS decoupling limit is dual to the Spin Matrix theory limit studied in \cite{Baiguera:2022pll} with  PSU$(1,2|3)$ as global symmetry. This Spin Matrix theory contains all other  Spin Matrix theories from $\CN=4$ SYM as subsectors.

An interesting feature of the geometry \eqref{met_Tdual_pq} for $p=q=2$ is that $\CP^2$ does not permit a spin structure, which means a massless particle supermultiplet is not allowed.
However, we are considering strings in a sector with a fixed non-zero winding number along $v$. Therefore, due to the coupling to the non-trivial Kalb-Ramond field \eqref{B_Tdual_pq}, such strings can admit supersymmetry \cite{Duff:1998us}.%
\footnote{It was pointed out in \cite{Duff:1997qz,Duff:1998us}
that T or S-dualities can relate a geometry which allows supersymmetry to a geometry that does not.} 
This is in line with the expectation that, upon 
including worldsheet fermions in the twisted BPS decoupling limits,
the isometry group SU$(1,2)\times \mbox{SU}(3)$ of the background \eqref{bg_su_pq} should be extended to PSU$(1,2|3)$, matching the global symmetry on the gauge theory side.

\section{Twisted limits of D-branes}
\label{sec:Dbranes}

In the previous section we have considered twisted BPS decoupling limits of closed strings on AdS$_5\times S^5$ in the global patch.
In this section we demonstrate that one can take these limits for D-branes as well. 
For D-branes on AdS$_5\times S^5$ the BPS states correspond to giant gravitons \cite{McGreevy:2000cw,Grisaru:2000zn,Hashimoto:2000zp}.
Indeed, we shall see below that the twisted BPS decoupling limit for a single D-brane corresponds to restricting the brane to a BPS configuration. 
In contrast, for multiple D-branes, one can use the non-abelian DBI action \cite{Taylor:1999pr,Myers:1999ps} to explore fluctuations near a BPS configuration.

Specifically, we consider twisted BPS decoupling limits of D-branes with $p=0$ corresponding to the BPS states with $E=\sum_{i=1}^{q+1} J_i$. We shall see that the geometry of the transverse space that the D-branes explore in these limits will be the same as for the string.

\subsection{Single D-brane}
\label{sec:single_Dbrane}

We consider in the following the 
twisted BPS decoupling limits with $p=0$ corresponding to BPS states with $E= w$ with
\begin{equation}
w = \sum_{i=1}^{q+1} J_i \,.
\end{equation}
As explained in Appendix \ref{app:null_world}, the corresponding null worldline sits in an $\mathbb{R}\times S^{2q+1}$ subset of AdS$_5\times S^5$. 
However, due to the Myers effect \cite{Myers:1999ps}, the D-brane BPS configurations expand beyond the subspace. 
Below we consider the specific scenario in which D3-branes are wrapped on an $S^3 \subset \mbox{AdS}_5$. These BPS states correspond to dual giant gravitons \cite{Grisaru:2000zn,Hashimoto:2000zp}.

To start, we consider the AdS$_5\times S^{2q+1}$ subset of AdS$_5\times S^5$ with metric
\begin{eqnarray}
\label{metric_gen_s5}
\frac{ds^2}{\alpha'} &=& R^2 \left[ - \cosh^2 \! \rho  \frac{dT^2}{\epsilon^4}
+ d\rho^2+ \sinh^2 \! \rho \, d\widetilde{\Omega}_3^2 
\right.
\nn \\ &&
+ \left(\frac{dT}{\epsilon^2} + d\hat{\gamma}+A^{(q)} \right)^2 
+ d\Sigma_q^{\, 2}
\Big]  \,.
\end{eqnarray}
In terms of this, the null worldline is at $\rho=0$, parametrized by $T$, with all other coordinates fixed. A dual giant graviton corresponds to a single D3-brane wrapped on the $S^3 \subset \mbox{AdS}_5$ at fixed $\rho > 0$, again with the path parametrized by $T$ and with all other coordinates fixed. This has $p_{\hat{\gamma}} = w$.

Now perform a T-duality along $\hat{\gamma}$.  
This gives the AdS$_5\times \CP^q \times S^1$ background with metric
\begin{eqnarray}
\label{metric_D4}
\frac{ds^2}{\alpha'} &=& R^2 \left[ - \cosh^2 \! \rho  \frac{dT^2}{\epsilon^4}
+ d\rho^2+ \sinh^2 \! \rho \, d\widetilde{\Omega}_3^2 
\right.
\nn \\ &&
+ d\Sigma_q^{\, 2}
\Big] +  \frac{dv^2}{R^2} \,,
\end{eqnarray}
and the Kalb-Ramond field and string coupling are
\begin{equation}
\label{B_D4}
\frac{B}{\alpha'} =- \left( \frac{dT}{\epsilon^2} + A^{(q)} \right)\wedge dv 
\spa
g_s = \frac{R^3}{4\pi N} \,.
\end{equation}
The Ramond-Ramond background fluxes are important as well.
Restricting to the subspace, one can express them through the six-form field strength
\begin{equation}
F_{(6)} = 4 \alpha'{}^{5/2} R^3 \sinh^3 \!\rho \cosh \rho \, dv \, dT \, d\rho \, \widetilde{\Omega}_3 \,.
\end{equation}
This corresponds to the five-form Ramond-Ramond potential
\begin{equation}
C_{(5)} =  \alpha'{}^{5/2} R^3 \sinh^4 \! \rho \, dv \, dT \, \widetilde{\Omega}_3 \,.
\end{equation}

A D3-brane wrapped on $S^3 \subset \mbox{AdS}_5$ and with momentum $p_{\hat{\gamma}}= w$ along $\hat{\gamma}$ corresponds after T-duality to a D4-brane wrapped on $S^3 \times S^1$, with the $S^1$ parametrized by $v$.
In line with this, we consider the following gauge choice for the D4-brane
\begin{equation}
\label{Dbrane_gauge}
T = \sigma^0 \spa  \chi_1 = \sigma^1 \spa \chi_2 = \sigma^2 \spa \chi_3 = \sigma^3 \spa v = \sigma^4 \,,
\end{equation}
where $\sigma^\alpha$, $\alpha=0,1,...,4$, are the worldvolume coordinates. We have the D4-brane Lagrangian
\begin{eqnarray}
\label{D4_L}
L&=& T_{\rm D4}  \int_{S^3 \times S^1} d^4 \sigma \left(  - \sqrt{-\det (\gamma_{\alpha\beta} + B_{\alpha\beta} ) } + C_{01234} \right)
\nn \\ 
&=& N \left[ - \frac{\sinh^3 \! \rho}{\alpha'}\sqrt{-\gamma_{TT} \gamma_{vv} - B_{Tv}^2  } +  \frac{\sinh^4 \! \rho}{\epsilon^2}  \right] \,,
\end{eqnarray}
where 
$d^4 \sigma = d\sigma^1 \cdots d\sigma^4$.
Here the tension is
\begin{equation}
T_{\rm D4} = \frac{1}{(2\pi)^4 g_s \alpha'{}^{5/2}}
= \frac{N}{4\pi^3 R^3 \alpha'{}^{5/2}} \,.
 \end{equation}
The momentum $p_{\hat{\gamma}}=w$ before T-duality is exchanged with electric flux  along $v$ given by
\begin{equation}
w= -\frac{\partial L}{\partial B_{Tv}}
=-   \frac{N \alpha' \sinh^3 \! \rho \, B_{Tv}}{\sqrt{
- \gamma_{TT} \gamma_{vv} - B_{Tv}^2 
 }} \,.
\end{equation}
For small $\epsilon$ we expand
\begin{eqnarray}
\label{sqrroot_exp}
&& \frac{1}{\alpha'}\sqrt{-\gamma_{TT} \gamma_{vv} - B_{Tv}^2  } 
\nn \\ &&= \frac{1}{\epsilon^2} \sqrt{\cosh^2 \rho - (1+ \epsilon^2 A_0^{(q)} )^2 } + \CO(\epsilon^2) \,.
\end{eqnarray}
Thus, for $\epsilon \rightarrow 0$ the electric flux is
\begin{equation}
\label{rho_fixed}
w = N \sinh^2 \rho \,.
\end{equation}
Therefore, the  $\epsilon \rightarrow 0$ limit of the Lagrangian is
\begin{equation}
\label{Dbrane_kin}
L = w \, A^{(q)}_i \dot{x}^i \,,
\end{equation}
with $\dot{x}^i = \partial_{0} x^i$.
This is identical to the kinetic term of the Lagrangian of the string in the same $p=0$ twisted BPS decoupling limit. This is seen from \eqref{galilean_action_fixed} by inserting the background \eqref{bg_su_pq} for $p=0$.
Thus, in both cases the kinetic term \eqref{Dbrane_kin} makes the $\CP^q$ target space a symplectic geometry.
This means the D-brane probes the same transverse symplectic geometry as the string.  

The $\epsilon\rightarrow 0$ limit fixes $\rho$ to be given by \eqref{rho_fixed} even if $\rho$ is dynamical before the limit. Let us consider the consistency of this. 
Firstly, we note that the leading term with a $\dot{\rho}$ coupling goes like $\epsilon^4$ relative to the leading behavior of \eqref{sqrroot_exp} and hence it vanishes in the limit. Therefore, it is consistent that $\rho$ does not depend on $\sigma^0$ after the limit. 
One can also consider allowing $\rho$ to depend on $\sigma^i$ with $i=1,2,3,4$.
This would give an additional term $\alpha' R^2 \partial_i \rho \partial_j \rho$ in the induced metric $\gamma_{ij}$ which would contribute at the leading order. However, this would mean that the $1/\epsilon^2$ divergent terms in \eqref{D4_L} would no longer cancel, corresponding to an infinitely steep confining potential for such modes. Therefore, in the twisted BPS decoupling limit $\epsilon\rightarrow 0$, only the modes with $\rho$ constant survive.

\subsection{Alternative formulation}
\label{sec:alt_form}

Above we have shown that in the twisted BPS decoupling limit the D-brane dynamics reduces to the Lagrangian \eqref{Dbrane_kin} on $\CP^q$, corresponding to dual giant graviton configurations moving on $S^5$. 
Below we show that one can alternatively formulate the dynamics in terms of harmonic oscillator dynamics on $\mathbb{C}^{q+1}$, connecting to previous results in the literature \cite{Kinney:2005ej,Biswas:2006tj,Mandal:2006tk} and \cite{Harmark:2016cjq}.

Consider the Lagrangian on $\mathbb{C}^{q+1}$
\begin{equation}
\label{L_C}
\tilde{L} = \sum_{k=1}^{q+1} \left( i z_k \dot{\bar{z}}_k -i \bar{z}_k \dot{z}_k - |z_k|^2 \right) \,.
\end{equation}
Write $z_k = u \, \eta_k$ where
\begin{equation}
\label{eta_k}
\eta_i = \frac{Z_i}{\sqrt{1+|Z|^2}} \spa \eta_{q+1}=  \frac{1}{\sqrt{1+|Z|^2}} \,,
\end{equation}
with $i=1,...,q$ and $|Z|^2=\sum_{i=1}^q |Z_i|^2$.
Then $Z_i$ parametrize $\CP^q$.
We compute
\begin{equation}
\tilde{L} =  i u \dot{\bar{u}} -i \bar{u} \dot{u} - |u|^2 
+ |u|^2 \sum_{k=1}^{q+1} \Big( i \eta_k \dot{\bar{\eta}}_k -i \bar{\eta}_k \dot{\eta}_k \Big) \,,
\end{equation}
using $\sum_{k=1}^{q+1} |\eta_k|^2 = 1$. 
We get $p_u= - i \bar{u}$ and $p_{\eta_k}=-i|u^2|^2 \bar{\eta}_k$.
Thus, $\tilde{H}=|u|^2$ which means $|u|$ is constant. Since $\tilde{H}=w$ one finds $|u| = \sqrt{w} = \sqrt{N} \sinh \rho$.
Using this, the dynamics reduces to 
\begin{equation}
\label{L_eta}
L =  w \sum_{k=1}^{q+1} \Big( i \eta_k \dot{\bar{\eta}}_k -i \bar{\eta}_k \dot{\eta}_k \Big) \,,
\end{equation}
which is equivalent to \eqref{Dbrane_kin}.%
\footnote{Note that the parametrization \eqref{eta_k} corresponds to a specific choice of overall phase of $\eta_k$
for which $\eta_{q+1}$ is real. Instead, one can freely choose the overall phase of $\eta_k$ to fit the parametrization of $A^{(q)}$.}

A useful alternative formulation is now found by performing the canonical transformation 
\begin{equation}
\label{can_trans}
\begin{array}{c} \ds
z_k = \frac{P_k + X'_k + i (X_k +  P'_k)}{\sqrt{2}} \,,
\\[4mm] \ds
p_{z_k} = \frac{-X_k + P'_k - i (P_k -  X'_k)}{\sqrt{2}}  \,,
\end{array}
\end{equation}
to the real phase-space variables $(X_k,X'_k,P_k,P'_k)$. 
From \eqref{L_C} one finds $p_{z_k}= -i \bar{z}_k$ and $p_{\bar{z}_k} = i z_k$ as constraints on the phase space. This is solved by $X'_k=P'_k=0$. Thus, we get an equivalent formulation of \eqref{L_C} in terms of the unconstrained phase-space variables $(X_k,P_k)$ with the Hamiltonian
\begin{equation}
\label{hatH}
\hat{H} = \frac{1}{2} \sum_{k=1}^{q+1} ( P_k^2 + X_k^2 ) \,,
\end{equation}
corresponding to $(q+1)$ one-dimensional harmonic oscillators of equal masses. 
This Hamiltonian connects with the observation that giant gravitons with $E=\sum_{i=1}^{q+1} J_i$ are associated with harmonic oscillators \cite{Kinney:2005ej,Biswas:2006tj,Mandal:2006tk}.
Moreover, it is exactly the Hamiltonian that emerges as
 the non-interacting part of the semi-classical regime of the Spin Matrix theory limits of $\CN=4$ SYM \cite{Harmark:2016cjq}.
 Note that \eqref{hatH} is conserved with $\hat{H} =  w$.

\subsection{Multiple D-branes}
\label{sec:multiple_Dbranes}

A single D-brane in a twisted BPS decoupling limit has no interaction term, as the only surviving modes are BPS states. 
However, below we shall see that multiple D-branes have non-BPS excitations in the twisted BPS decoupling limit. Multiple D-branes are described by the non-abelian DBI action \cite{Taylor:1999pr,Myers:1999ps} (although the description is not complete). 

In the following, we consider $k$ D4-branes on the T-dual background \eqref{metric_D4}-\eqref{B_D4}, wrapped on the same $S^3\times S^1$ subspace as in the single D4-brane case.
We work in the probe limit $k \ll N$. 

In the following calculation
we set the non-abelian worldvolume gauge field  to zero since a non-zero field strength would affect the cancellation of the divergence in the twisted BPS decoupling limit. Note, however, that one should take into account the Gauss constraint that arises from this. 
As before, we choose the gauge \eqref{Dbrane_gauge}. Moreover, we assume that the transverse coordinates can only depend on worldvolume time, since any dependence on the $S^3\times S^1$ worldvolume directions would spoil the cancellation of the $1/\epsilon^2$ divergence.
Under these assumptions, the Lagrangian is
\begin{equation}
L = T_{\rm D4}  \int_{S^3 \times S^1} d^4 \sigma \, \mbox{STr} \left(  - S_1 S_2 + C_{01234} \right) \,,
\end{equation}
where we defined
\begin{equation}
\begin{array}{c} \ds
S_1 = \sqrt{-\det (\gamma_{\alpha\beta} + B_{\alpha\beta} + M_{\alpha\beta} 
) }
\spa S_2 = \sqrt{\det Q^I_J  } \,,
\\[4mm] 
M_{\alpha\beta}=\partial_\alpha x^I g_{IJ}  \Big((Q^{-1}) ^{J}{}_{K} - \delta^{J}_{K} \Big) \partial_\beta x^K \,,
\\[4mm] 
Q^I{}_J = \delta^I_J + \frac{1}{\alpha'}F^{IK} g_{KJ} 
\spa
F^{IJ} = \frac{i}{2\pi} [x^I , x^J ]  \,,
\end{array}
\end{equation}
where $I,J,K,L$ run over the transverse directions of the D4-branes and $g_{IJ}$ are metric components in those directions without pullback. Moreover, $\mbox{STr}$ is the symmetric trace. As is well-known, the accuracy of the non-abelian DBI action is unclear beyond terms that are fourth order in the gauge field strength, here counting either $\partial_\alpha x^I$ or $F^{IJ}$. However, this will not be an issue for our calculation.

In the twisted BPS decoupling limit $R^2=\epsilon\rightarrow 0$ the  $M_{00}$ contribution in $S_1$ vanishes since it goes like $\epsilon^3$ and the  leading contribution from $\gamma_{00}$ goes like $\epsilon^{-3}$. Thus, $S_1$ gives the matrix-valued equivalent of \eqref{Dbrane_kin} plus the divergent term that cancels with $C_{01234}$.
Instead, $S_2$ for small $R^2=\epsilon$ is expanded as
\begin{equation}
\mathbb{I} - \frac{1}{4\alpha'{}^2} g_{IJ} F^{JK} g_{KL} F^{LI} + \CO(\epsilon^4) \,,
\end{equation}
where $\mathbb{I}$ is the identity matrix. Since the subleading term is of order $\epsilon^2$ relative to the leading divergence, it survives the twisted BPS decoupling limit.

Therefore, the resulting Lagrangian describing multiple D-branes in the $\epsilon \rightarrow 0$ limit is 
\begin{equation}
\label{Dbrane_full}
L = \mbox{STr} \left( w \,  A^{(q)}_i  \dot{x}^i
+ \frac{w^2}{4N} h_{ij} F^{jk} h_{kl} F^{li}
\right) \,,
\end{equation}
where $h_{ij} dx^i dx^j = d\Sigma_q^{\, 2}$.
The Gauss constraint is
\begin{equation}
\label{gauss_Dbrane_full}
\sum_i [ x^i , A^{(q)}_i ] = 0 \,.
\end{equation}
We see from \eqref{Dbrane_full} that in addition to the matrix-valued version of the kinetic term \eqref{Dbrane_kin} we have an interaction term that couples to $h_{ij}$. This interaction term describes the surviving near-BPS excitations in the twisted BPS decoupling limit.
This shows that the non-BPS non-abelian D-brane excitations probe the same U$(1)$-Galilean background \eqref{bg_su_pq} (for $p=0$) as a fundamental string in the twisted BPS decoupling limit.

Both $A^{(q)}_i$ and $h_{ij}$ are non-linear functions of the coordinates describing $\CP^q$ which are now promoted to matrices. 
For this reason, the  Lagrangian \eqref{Dbrane_full} is complicated to work with. However, we can consider instead the alternative formulation of the D-branes explained above in Section \ref{sec:alt_form}.
In this formulation, one first reparametrizes $\CP^q$ in terms of $\eta_k$ in \eqref{eta_k}. This means the kinetic term should be the matrix-valued version of \eqref{L_eta}. Using $z_k = u \, \eta_k$ the dynamics is equivalent to the matrix-valued version of \eqref{L_C}. Finally, one performs the canonical transformation \eqref{can_trans} and solves the constraints $p_{z_k}= -i \bar{z}_k$ and $p_{\bar{z}_k} = i z_k$ by imposing $X'_k=P'_k=0$. Applying these steps to the interacting term in \eqref{Dbrane_full} as well, we get the equivalent Hamiltonian
\begin{equation}
\label{hat_H}
\hat{H} = \frac{1}{2} \sum_i \tr \! ( y_i^2 )
- \frac{1}{32\pi^2N} \sum_{i<j} \tr \! ([y_i , y_j]^2 )  \,,
\end{equation}
where we defined the $2q+2$ phase-space variables 
\begin{equation}
y_{2k-1} = X_k \spa y_{2k} = P_k \ \ \mbox{for} \ \ k=1,...,q+1 \,.
\end{equation}
In terms of these variables, the Gauss constraint becomes
\begin{equation}
\sum_{k=1}^{q+1} [ y_{2k-1} , y_{2k} ] = 0 \,.
\end{equation}
The Hamiltonian \eqref{hat_H} was previously derived for the $q=1$ case in \cite{Harmark:2016cjq} from a D3-brane point of view. That derivation is clarified and extended in the upcoming work \cite{futureSMT}. It would be interesting to explore the equivalence of this description to that of Eqs.~\eqref{Dbrane_full}-\eqref{gauss_Dbrane_full} further given the matrix-valued nature of the geometry.

As pointed out in \cite{Harmark:2016cjq}, the 
Hamiltonian \eqref{hat_H} matches exactly the 
semi-classical limit of the SU$(2)$ Spin Matrix theory for $q=1$, in the regime with $w \gg 1$ and $k$ being finite. Thus, just as in the string case, one can match the gauge theory and string theory sides in the twisted BPS decoupling limit. This is remarkable, as it is a match in the opposite regime of the planar limit considered in Section \ref{sec:examples}, while the gauge theory side is in a weak 't Hooft coupling regime \eqref{weak_thooft}.

\section{Conclusions and outlook}

In this work we have established a new class of BPS decoupling limits.
A twisted BPS decoupling limit 
is associated with a momentum carrying BPS state on 
 a geometrically twisted circle fiber, unlike a DLCQ limit for which the circle is untwisted. 
As for a DLCQ limit, the target space geometry becomes non-Lorentzian in the BPS limit. This is apparent from the T-dual frame, in which momentum is replaced with winding. For a single string probe, one obtains the sigma-model \eqref{galilean_action} describing the BPS and near-BPS excitations
with a symplectic U$(1)$-Galilean target space.
We have exhibited in detail the possible BPS decoupling limits
for single string probes on AdS$_5\times S^5$ in the global patch. Here the string lives in a symplectic K\" ahler geometry after the limit. Moreover, we have shown that one can take twisted BPS decoupling limits of D-branes on AdS$_5\times S^5$ as well. This shows that D-branes probe the same symplectic geometry as the string. 



Do the twisted BPS decoupling limits encompass all possible BPS decoupling limits of type IIB string theory on AdS$_5\times S^5$ in the global patch?
Yes, this is actually the case:
\begin{itemize}
\item Strictly speaking, there is one exception to this statement: the 
$p=q=0$ limit toward a $1/2$ BPS state mentioned in Section \ref{sec:examples}. Here the transverse space is a point, and hence no twist is involved. All other BPS decoupling limits involve a twist.
\item The statement assumes that no particle or brane charge is turned on, as that would give a richer spectrum of BPS states. 
As an example, think of the $1/4$ BPS states with a perpendicular D3-brane and D-string, or a perpendicular D3-brane and F-string. In the Poincar\'e patch of AdS$_5\times S^5$ these cases correspond to turning on magnetic or electric particle charges in the dual $\CN=4$ SYM. The BPS decoupling limits related to these BPS states are explored in \cite{Lambert:2024yjk,Blair:2024aqz,Lambert:2024ncn,Harmark:2025ikv,Kim:2026hpe}. Note that one cannot add such particle charges to obtain new BPS states in the global patch, since in this case the gauge theory side is on a three-sphere which does not allow for a net charge.  
\item
One could consider a DLCQ-like decoupling limit of 
type IIB string theory on AdS$_5\times S^5$ in the global patch.%
\footnote{A similar limit on the dual gauge theory side is considered in \cite{Baiguera:2024vlj}. In this work
the null-reduction of $\CN=4$ SYM on $\R \times S^3$ is analyzed, and shown to be related by a conformal mapping to $\CN=4$ SYM on Minkowski space with a $\Omega$-deformation.}
However, as mentioned in Sections \ref{sec:twistedlimit} and \ref{sec:examples}, there is no BPS state that would correspond to such a limit in the global PSU$(2,2|4)$ algebra. Hence it is unlikely to yield a consistent quantum theory of strings. However, combined with a second limit as part of the two-step limiting procedure of \cite{Harmark:2017rpg,Harmark:2018cdl,Harmark:2019upf,Harmark:2020vll,Bidussi:2023rfs} it reproduces the twisted BPS decoupling limits of this paper.
\item
Non-relativistic string limits of type IIB string theory on AdS$_5\times S^5$ in the global patch have also been investigated \cite{Gomis:2005pg,Fontanella:2021hcb,Fontanella:2024rvn,Fontanella:2024kyl}. 
Such limits would require identifying a circle direction that the string has a winding mode, along with a critical Kalb-Ramond field. 
However, both of these requirements take us outside the AdS/CFT correspondence, as they change the global boundary conditions. 
\end{itemize}

{\bf Future directions:}
The twisted BPS decoupling limits of type IIB string theory on AdS$_5\times S^5$ zoom in on the dynamics close to BPS states within the AdS/CFT correspondence. As already hinted by the gauge theory side \cite{Harmark:2014mpa,Harmark:2019zkn,Baiguera:2020jgy,Baiguera:2020mgk,Baiguera:2021hky,Baiguera:2022pll}, this allows for a more tractable effective theory close to BPS. Therefore, this work opens up several new directions: 
\begin{itemize}
\item An important direction would be to formulate the fundamental string for finite $w$ in the twisted BPS decoupling limits of
AdS$_5\times S^5$.
Since the decoupling limit should lead to a simplification, this could potentially show a path toward a proof of holography in this simplified and more tractable setting. In turn, it could also reveal a conformal field theory formulation of the sigma-model.
\item
Turning on the string coupling, strings can split or join. A full theoretical understanding of this is still lacking. However, for twisted BPS decoupling limits of 
AdS$_5\times S^5$ there are several reasons to expect a significant simplification, which could lead to a tractable theory. 
For one thing, the multi-string states in the limits  should correspond to partitions of the winding number, similarly to what happens in non-relativistic string theory and matrix string theory \cite{Motl:1997th,Dijkgraaf:1997vv,Gomis:2000bd,Danielsson:2000gi,Harmark:2025ikv}. 
Moreover, on the holographically dual side, it is clear that the Spin Matrix theory limit reveals a simpler structure, even non-perturbatively for finite string coupling. 
This could possibly provide a route to understanding the second quantization of strings in this more tractable setting. 
\item
One could turn the twisted decoupling limit of strings on AdS$_5\times S^5$ into an expansion in powers of $R^4$, which is proportional to the 't Hooft coupling. 
While the strict limit captures the one-loop term in $\CN=4$ SYM, the expansion could capture the higher loops as well. 
This could reveal new geometric structures, in addition to the symplectic U$(1)$-Galilean structure in the strict limit. 
To this end, one can use the techniques developed in \cite{Hansen:2019pkl,Hartong:2021ekg} on expansions in gravity and string theory. 
\item 
In the maximal limit $p=q=2$, 
one could implement a twisted BPS decoupling limit of the type IIB supergravity action, similarly to how one takes DLCQ-related limits of supergravity actions \cite{Bergshoeff:2021bmc,Blair:2021waq,Bergshoeff:2024ipq,Bergshoeff:2025grj,Maskalaniec:2026vlk}. This could reveal the non-Lorentzian gravitational dynamics of U$(1)$-Galilean  
geometry relevant to both fundamental strings and D-branes. 
Potentially, this would enable one to see 
how gravitational dynamics emerge from a microscopic theory, since for the AdS/CFT limits  the dual Spin Matrix theory should capture the gravitational dynamics. 
\item 
For the maximal case $p=q=2$ of Section \ref{sec:examples}, the BPS states include not only configurations with strings and D-branes, but also a supersymmetric black hole \cite{Kunduri:2006ek}. 
Hence, one can explore the manifestations of this black hole in the twisted BPS decoupling limit, both on the string and gauge theory sides of the AdS/CFT correspondence. For instance, does it have a description in terms of a non-Lorentzian gravitational  theory?
\item
The U-duality orbit of DLCQ limits includes various BPS decoupling limits with holographic correspondences. 
Thus, exploring the U-duality orbit of twisted BPS decoupling limits could potentially give rise to new connections between holographically dual theories.
\item For the $p=1,2$ limits of AdS$_5\times S^5$, the U$(1)$-Galilean geometries include the complex hyperbolic space $\mCH^p$ which has a boundary. It would be interesting to examine what role this boundary plays in
 formulating a holographic correspondence between string theory and the dual Spin Matrix theory. It could possibly be an analogue of the conformal boundary of AdS$_5$.
More generally, one could further explore the holographic dualities between Spin Matrix theories and twisted BPS decoupling limits on AdS$_5\times S^5$. Related work includes \cite{Harmark:2019zkn,Baiguera:2020jgy,Baiguera:2020mgk,Baiguera:2021hky,Baiguera:2022pll,Baiguera:2024vlj,Auzzi:2026gan}.
\item
Turning to twisted BPS decoupling limits for D-branes in AdS$_5\times S^5$, we considered in this paper the cases with $p=0$. However, it is well known that there are  
BPS configurations with $p=1,2$ as well \cite{Kim:2006he}.
Another possible direction is to consider twisted BPS decoupling limits of open strings ending on D-branes.
\item
In general, a full non-abelian DBI action describing multiple D-branes is still lacking. However, in the twisted BPS decoupling limit, these limitations might not apply. This would be interesting to explore, also since it opens a window into a matrix formulation of target space geometry of strings and D-branes in the finite string coupling regime.
\item
In this work, we have focused on twisted BPS decoupling limits of AdS$_5\times S^5$. 
However, there are several other backgrounds with geometrically twisted circle fibers for which one can perform twisted BPS decoupling limits.
These include holographic backgrounds such as
AdS$_4\times \CP^3$ or AdS$_3\times S^3$. 
\end{itemize}

\section*{Acknowledgements}

We warmly thank Stefano Baiguera, Jelle Hartong, Neil Lambert, Yang Lei, Carlos Nunez, Niels Obers and Ziqi Yan for great discussions and valuable comments on the manuscript. We acknowledge support from the “Center of Gravity”, a Center of Excellence funded by the Danish
National Research Foundation under Grant N. DNRF184.


\begin{appendix}
\section{Null worldlines for AdS$_5\times S^5$ BPS states}
\label{app:null_world}

In this appendix we identify the null worldlines, and the associated subspaces of AdS$_5\times S^5$ in which they can be embedded, for each of the BPS states \eqref{BPS_form} of PSU$(2,2|4)$. 
We use the metric of AdS$_5\times S^5$ in the global patch written as \eqref{gen_adss5}.

The BPS states \eqref{BPS_form} are labeled by $p\geq 0$ and $q\geq 0$ with $p=q=0$ corresponding to $1/2$ BPS, $p+q=1$ to $1/4$ BPS, $p+q=2$ to $1/8$ BPS and $p=q=2$ to $1/16$ BPS.

{\bf Null worldlines for $p=0$:}
The BPS states \eqref{BPS_form} with $p=0$ and $q=0,1,2$ correspond to null momentum states moving on $S^5$. Thus, they have $\rho=0$ in \eqref{gen_adss5} corresponding to $\mathbb{R} \subset \mbox{AdS}_5$. We now describe the null worldline for each case $q=0,1,2$.

For $q=0$ we write the $S^5$ metric as
\begin{equation}
d\Omega_5^{\, 2} = d\xi^2 +  \cos^2 \! \xi \, d\gamma^2 + \sin^2 \! \xi \, d\Omega_3^{\, 2} \,.
\end{equation}
In terms of this, the $(p,q)=(0,0)$ null worldline has $\gamma-t$ fixed, $\rho=0$ and $\xi=0$, which we can embed in the subspace $\mathbb{R} \times S^1\subset \mbox{AdS}_5\times S^5$.

For $q=1$ we use the parametrization of $S^5$ recorded in \eqref{metric_fullsu2} and \eqref{metpot_S3}. Thus, the $(p,q)=(0,1)$ null worldline is defined by $\rho=0$, $\xi=0$, $\gamma-t$ constant and the angles of $d\Sigma_1^{\, 2}$ fixed. Therefore, as already reviewed in Section \ref{sec:twistedlimit}, we can embed this null worldline in the subspace $\mathbb{R} \times S^{3} \subset \mbox{AdS}_5\times S^5$. 
Note that below we use the notation $A^{(1)} = A = \tfrac{1}{2} \cos \theta \, d\phi$ where the $(1)$ refers to $q=1$.

For $q=2$ we write the metric of $S^5$ as
\begin{equation}
d\Omega_5^{\, 2} = d\Sigma_2^{\, 2} + (d\gamma + A^{(2)})^2  \,,
\end{equation}
with
\begin{eqnarray}
&d\Sigma_2^{\, 2}= d\xi^2 + \cos^2 \xi \, d\Sigma_{1}^{\, 2}
 +\sin^2 \xi \cos ^2 \xi ( d\eta + A^{(1)}  )^2 \,,
 & 
\nn \\[2mm] 
& A^{(2)} = - \frac{2}{3} d\eta + \cos^2 \xi (d\eta + A^{(1)} ) \,.
&
\end{eqnarray}
The $(p,q)=(0,2)$ null worldline is now defined by $\rho=0$, $\xi=0$, $\gamma-t$ constant and the angles of $d\Sigma_1^{\, 2}$ fixed. This is embedded in the $\mathbb{R} \times S^{5}$ subspace of $\mbox{AdS}_5\times S^5$.

Thus, for all three cases $q=0,1,2$ we can write the $\mathbb{R} \times S^{2q+1}$ subspace metric as
\begin{equation}
\label{RxSq}
\frac{ds^2}{\alpha' R^2} = -dt^2 +(d\gamma+A^{(q)})^2 
+ d\Sigma_q^{\, 2} \,,
\end{equation}
with the understanding that $A^{(0)}=0$ and $d\Sigma_0^{\, 2}=0$. 
Therefore, in all three cases we follow the BPS momentum state by performing the coordinate transformation 
$t=T/\epsilon^2$ and $\gamma=\hat{\gamma}+T/\epsilon^2$, which means $E_T = (E- \sum_{i=1}^{q+1} J_i)/\epsilon^2$, $\hat{\gamma}$ is periodic with period $2\pi$ and $-i\partial_{\hat{\gamma}} = \sum_{i=1}^{q+1} J_i$. This gives the metric for $\mathbb{R} \times S^{2q+1}$ 
\begin{equation}
\label{subspace_p0}
\frac{ds^2}{\alpha' R^2} = -\frac{dT^2}{\epsilon^4} +\left(\frac{dT}{\epsilon^2} +d\hat{\gamma}+A^{(q)}\right)^2 
+ d\Sigma_q^{\, 2} \,,
\end{equation}
where the null worldline is parametrized by $T$ with all other coordinates held fixed. This corresponds to \eqref{subspace_pq} for $p=0$.

{\bf Null worldlines with $p=1$:} For the BPS states with $p=1$ and $q=0,1$ we should include a momentum corresponding to the angular momentum $S_1$ of AdS$_5$. To this end 
we parametrize the three-sphere within AdS$_5$ as 
\begin{equation}
\label{ads_s3a}
d\widetilde{\Omega}_3^{\, 2}= d\zeta^2 + \cos^2 \! \zeta \, d\phi_1^2
+ \sin^2 \! \zeta \, d\phi_2^2  \,,
\end{equation}
with $S_j=-i \partial_{\phi_j}$. The BPS state $E = S_1+  \sum_{i=1}^{q+1} J_i$ moves in particular along $\phi_1$ with $\zeta=0$. Thus, it has a null worldline in an AdS$_3 \times S^{2q+1}$ subspace with metric
\begin{eqnarray}
\label{ads_Sq_sub}
\frac{ds^2}{\alpha' R^2} 
&=& - \cosh^2 \! \rho \, dt^2 + d\rho^2 + \sinh^2 \! \rho \, d\phi_1^2 
\nn \\ &&
+ (d\gamma+A^{(q)})^2 
+ d\Sigma_q^{\, 2} \,,
\end{eqnarray}
where we used the above results and notation on the $\mathbb{R} \times S^{2q+1}$ cases for the $S^5$ part. In terms of this, the null worldline has $\phi_1-t$ and $\gamma-t$ constant, while $\rho$ and the angles of $d\Sigma_q^{\, 2}$ are fixed.
Therefore, we can follow the BPS momentum state by 
making the coordinate transformation
\begin{equation}
t = \frac{T}{\epsilon^2} \spa \phi_1 = \hat{\phi}_1 + \frac{T}{\epsilon^2} \spa \gamma = \hat{\gamma} + \frac{T}{\epsilon^2} \,,
\end{equation}
which gives the 
AdS$_3 \times S^{2q+1}$ metric as
\begin{eqnarray}
\label{subspace_p1}
\frac{ds^2}{\alpha'  R^2} &=& - \left( \frac{dT}{\epsilon^2}     - \tilde{A}^{(1)} \right)^2  + \left(\frac{dT}{\epsilon^2} + d\hat{\gamma} + A^{(q)} \right)^2  
\nn \\[2mm] &&
+ dH_1^{\, 2}+  d\Sigma_q^{\, 2} \,,
\end{eqnarray}
where 
\begin{equation}
\label{met_CH1}
dH_1^{\, 2} = d\rho^2 + \sinh^2 \! \rho \cosh^2 \! \rho \, d\hat{\phi}_1^2 \,,
\end{equation}
is the metric of one-dimensional complex hyperbolic space $\mCH^1$
and we defined the one-form connection
\begin{equation}
\label{pot_CH1}
\tilde{A}^{(1)} = \sinh^2 \! \rho \,   d\hat{\phi}_1 \,.
\end{equation}
Thus, in terms of the metric \eqref{subspace_p1} the null worldline is parametrized by $T$ with all other coordinates held fixed.
The metric \eqref{subspace_p1} corresponds to \eqref{subspace_pq} for $p=1$.

{\bf Null worldline for $p=2$:}
Finally, for the BPS states with $p=2$ and $q=0,2$ we turn on the angular momentum combination $S_1+S_2$ on AdS$_5$. Therefore, we write the metric of the three-sphere within AdS$_5$ as
\begin{equation}
d\widetilde{\Omega}_3^{\, 2}
=d\widetilde{\Sigma}_1^{\, 2} 
+   (d\psi+ V)^2  \,,
\end{equation}
where
\begin{equation}
d\widetilde{\Sigma}_{1}^{\, 2} = \frac{1}{4} ( d\vartheta^2 + \sin^2 \vartheta\, d\varphi^2 )
\spa
V = \frac{1}{2} \cos \vartheta \, d\varphi \,,
\end{equation}
so that $S_1+S_2 = -i \partial_\psi$.
Using again the results and notation on the $\mathbb{R} \times S^{2q+1}$ cases for the $S^5$ part we can follow the BPS momentum state \eqref{BPS_form} with $p=2$ and $q=0,2$ using the coordinate transformation
\begin{equation}
t = \frac{T}{\epsilon^2} \spa \psi = \hat{\psi} + \frac{T}{\epsilon^2} \spa \gamma = \hat{\gamma} + \frac{T}{\epsilon^2} \,,
\end{equation}
which gives
\begin{eqnarray}
\label{subspace_p2}
\frac{ds^2}{\alpha'  R^2} &=& - \left( \frac{dT}{\epsilon^2}     - \tilde{A}^{(2)} \right)^2  + \left(\frac{dT}{\epsilon^2} + d\hat{\gamma} + A^{(q)} \right)^2  
\nn \\[2mm] &&
+ dH_2^{\, 2}+  d\Sigma_q^{\, 2} \,.
\end{eqnarray}
Here 
\begin{equation}
dH_2^{\, 2} =  d\rho^2
+ \sinh^2 \! \rho \, d\widetilde{\Sigma}_1^{\, 2} + \sinh^2 \! \rho \cosh^2 \! \rho \, (d\hat{\psi} + V)^2  \,,
\end{equation}
is the metric of two-dimensional complex hyperbolic space $\mCH^2$ and we have introduced the one-form
\begin{equation}
\tilde{A}^{(2)} =  \sinh^2 \! \rho (d\hat{\psi} + V)  \,.
\end{equation}
In terms of the metric \eqref{subspace_p2} the null worldline is parametrized by $T$ with all other coordinates held fixed.
The metric \eqref{subspace_p2} corresponds to \eqref{subspace_pq} for $p=2$.

\end{appendix}

\providecommand{\href}[2]{#2}\begingroup\raggedright\endgroup

\end{document}